\documentclass[12pt,a4paper]{article}

\usepackage{fullpage}
\usepackage{blkarray}
\usepackage[applemac]{inputenc}
\usepackage{graphicx}
\usepackage{enumerate}
\usepackage{color}
\usepackage{natbib}
\usepackage{amsfonts}
\usepackage{amssymb,amsmath}
\usepackage{sidecap}
\usepackage{subfig}
\usepackage{float}
\usepackage[table]{xcolor}
\usepackage{nicefrac}
\usepackage{ulem}
\usepackage{tikz} 
\usepackage{bbm}
\usepackage{authblk}

\usepackage{tikz}
\usepackage{float}
\usetikzlibrary{shapes,backgrounds}
\usetikzlibrary{patterns}
\usetikzlibrary{positioning}
\usetikzlibrary{graphs,graphs.standard}
\usetikzlibrary{arrows.meta,bending,quotes}

\usepackage{xcolor}
\colorlet{myblue}{blue}
\colorlet{mygreen}{green!60!black}
\colorlet{myred}{red!60!black}
\colorlet{myred2}{red!80!black}

\tikzstyle{node in}=[thin,circle,minimum size=0.5cm,draw=black,fill=mygreen!25]
\tikzstyle{node out}=[thin,circle,minimum size=0.5cm,draw=black,fill=myred2!25]

\usepackage[colorlinks = true,
            linkcolor = blue,
            urlcolor  = blue,
            citecolor = blue,
            anchorcolor = blue]{hyperref}

\newtheorem{theorem}{Theorem}
{\end{theorem}\vskip.2cm}
\newtheorem{claim}{Claim}
\newenvironment{cl}{\vskip.2cm\begin{claim}}%
{\end{claim}\vskip.2cm}
\newtheorem{fact}{Fact}
{\end{fact}\vskip.2cm}
\newtheorem{lemma}{Lemma}
{\end{lemma}\vskip.2cm}
\newtheorem{corolla}{Corollary}
{\end{corolla}\vskip.2cm}
\newtheorem{defini}{Definition}
\newenvironment{defi}{\vskip.2cm\begin{defini}}%
{\end{defini}\vskip.2cm}
\newtheorem{proposi}{Proposition}
{\end{proposi}\vskip.1cm}
\newtheorem{prop}{Proposition}
{\end{prop}\vskip.1cm}
\newtheorem{cla}{Claim}
{\end{cla}\vskip.2cm}
\newtheorem{assump}{Assumption}
{\end{assump}\vskip.1cm}
\newtheorem{hypoth}{Assumption}

{\end{hypoth}\vskip.1cm}
\newtheorem{demo}{Proof}

{\cqfd \end{demo}}
\newtheorem{remark}{Remark}
{\end{remark}\vskip.3cm}
\newtheorem{example}{Example}
{\end{example}\vskip.3cm}

\def\1{\mathbf{1}}
\def\a{{\overline a}}

\def \0{{\mathbf{0}}}

\def\cqfd {$\Box$}
\def\co{{\rm co}\hskip 1pt}
\def \diag {\mathrm{diag}}

\def\E{{\mathcal E}}

\def\diag{{\rm diag}}

\def\M{{\mathcal M}}

\def\N{{\mathbb{N}}}

\def\p{\overline{p}}

\def\proj{{\rm proj}}
\def\proj0{{\rm proj}_{\co M_0}}

\def\R{{\mathbb{R}}}

\def\v{\overline{v}}

\def\x{\overline{x}}

\def\y{\overline{y}}

\makeatletter

\newcommand{\Rmnum}[1]{\expandafter\@slowromancap\romannumeral #1@}
\makeatother

\title{Strategic formation of production networks}

\author[1,2]{Antoine Mandel\thanks{\href{antoine.mandel@univ-paris1.fr}{antoine.mandel@univ-paris1.fr}}}
\author[3]{Van-Quy Nguyen\thanks{\href{quynv@neu.edu.vn}{quynv@neu.edu.vn}}}
\author[4]{Bach Dong-Xuan\thanks{\href{bach.dong@uni-bielefeld.de}{bach.dong@uni-bielefeld.de}}}
\affil[1]{\textit{\small Centre d'Economie de la Sorbonne, 106-112 Bd de l'H\^{o}pital, 75647 Paris, France}}
\affil[2]{\textit{\small Paris School of Economics, Universit\'{e} Paris 1 Panth\'{e}on-Sorbonne}}
\affil[3]{\textit{\small Faculty of Mathematical Economics, National Economics University, Vietnam}}
\affil[4]{\textit{\small Center for Mathematical Economics, Bielefeld University}}

\begin{document}

\makeatletter
\renewcommand{\@biblabel}[1]{}
\makeatother
\maketitle

\begin{abstract}
We provide a strategic model of the formation of production networks that subsumes the standard general equilibrium approach. The objective of firms in our setting is to choose their supply relationships so as to maximize their profit at the general equilibrium that unfolds.  We show that this objective is equivalent to the maximization  by the firms of their eigenvector centrality in the production network. As is common in network formation games based on centrality, there  are multiple Nash equilibria in our setting. We have investigated the characteristics and the social efficiency of these equilibria in a stylized version of our model representing international trade networks. We show that the impact of network structure on social welfare is firstly determined by a trade-off between costs of increasing process complexity and positive spillovers on productivity induced by the diversification of the input mix. We further analyze a variant of our model that accounts for the risks of disruption of supply relationships. In this setting, we characterize how social welfare depends on the structure of the production network, the spatial distribution of risks, and the process of shock aggregation in supply chains. We finally show that simple trade policies characterized by sets of links that are either prevented or catalyzed can be a powerful equilibrium selection device. 
 \end{abstract}


\vspace{2 mm}

\newpage

\section{Introduction}

Decentralized network formation processes can lead to substantial inefficiencies and welfare costs \citep{jackson1996strategic}. The recent literature has particularly emphasized that linkage behavior aligned with micro-level objectives can have substantial negative externalities such as polarization in social networks \citep{levy2021social}, epidemic spreading in human networks \citep{pastor2001epidemic,antras2023globalization}, the emergence of systemic risk in financial \citep{acemoglu2015systemic} or macro-economic networks \citep{acemoglu2012network,elliott2022supply}. In the productive realm, there is increasing political scrutiny of the negative externalities potentially associated to the decentralized formation of global supply chains and production networks, i.e. of globalization  \citep{witt2019globalization,kornprobst2021globalization}. Notably, globalization of supply chains has been identified as a potential source of increasing social inequality \citep{costinot2012global} and as a potential channel of propagation of global risks \citep{razin2020globalization,irwin2020pandemic}.

Existing  models of the formation of supply relationships are mostly silent about these potential negative externalities because they generally consider a setting where the incentives of firms and consumer are essentially aligned. Indeed, both in the canonical models of the international trade literature \citep{eaton2002technology,alvarez2007general}  as well as in the specific literature on the endogenous formation of production networks \citep{acemoglu2020endogenous}, firms seek to minimize production costs. This objective is aligned with utility maximization for a representative household. In this paper, we adopt a polar perspective whereby firms strategically choose their supply relationships, and thus their  production function,  with the aim of maximizing profits at the general equilibrium that will ensue. Hence our model, as this of  \citep{acemoglu2020endogenous}, subsumes the standard general equilibrium approach by considering choices that define the production structure of the economy rather than taking it as a primitive. In this setting that is beyond general equilibrium, cost minimization and profit maximization correspond to different incentives and can lead to very different outcomes. Indeed, profit and revenue maximization might be independent, or even antithetical, to cost and price minimization, depending e.g. on the price elasticity of demand. Accordingly, strategic behavior aiming at profit maximization may generate negative external effects on social welfare. Our aim in this paper is to analyze the potential scope of these inefficiencies, their drivers, as well as potential mitigation policies. 

We consider a setting similar to that of \citep{acemoglu2020endogenous} where firms' choices of supply relationships determine, at the micro level,  their productivity and the characteristics of their production function and, at the macro level, the productive structure of a general equilibrium economy. We assume that firms' objective is to maximize profits at the general equilibrium of the resulting economy. This defines a strategic network formation game. We focus on the Nash equilibrium of this game and its welfare properties. We first show the existence of such a Nash equilibrium, which is non-trivial because of non-convexities in the payoff function. We then provide a general characterization of equilibrium behavior and of social welfare in terms of network characteristics. Namely, we show that firms aim to maximize total incoming connectivity from the household, whereas social welfare depends on the connectivity from high-productivity firms towards the household. We investigate, in a specific class of replicate economies, the costs this can induce in terms of social welfare. Namely, we provide a closed-form expression for the price of anarchy and show that it can become arbitrarily large as certain Nash equilibria can be extremely inefficient in terms of aggregate productivity. We further investigate the interplay between strategic behavior and efficiency in a context where supply relationships can be disrupted by exogenous shocks such as natural hazards. That is, we investigate whether decentralized network formation processes can have negative externalities in terms of resilience of the network. As in the case of financial networks investigated in \cite{acemoglu2015systemic}, we find that the comparative resilience of the equilibrium network structure depends on the type of risks faced. When risks are ``small", diversified, highly connected, network structures are more resilient to shocks. When risks are large, prone to amplification, the benefits of diversification can be offset by increasing risk exposure, and less connected network structures are more resilient. Finally, we investigate policy measures that can nudge firms towards efficient equilibria by restricting the set of admissible supply relationships, e.g. through preferential trade agreements. Such measures can be very efficient: A few targeted interventions can substantially reduce the indeterminacy of equilibrium and the scope of potential inefficiencies.  

Our results are obtained in a highly stylized setting where all production functions are assumed to be Cobb-Douglas. In this specific case, firms are strategically ``neutral" about production costs because equilibrium revenues and profits are independent of productivity. This implies in particular that firms do not have incentives to minimize production costs as opposed to \cite{acemoglu2020endogenous}. Setting with alternative assumptions on elasticities of substitution could substantially modify our results. If intermediary inputs were perfect substitutes, firms would have incentives to target minimal production costs during the network formation step in order to increase their revenues in the resulting general equilibrium. Oppositely, if intermediary inputs were perfect complements, firms would have incentives to target maximal production costs during the network formation step in order to increase their revenues in the resulting general equilibrium.  

Our analysis can be put in perspective through the lenses of network theory and of the macroeconomic analysis of production networks. From the point of view of network theory, we show that the profit maximization objective of firms can equivalently be expressed as the maximization of their eigenvector centrality in the production network. In such centrality maximization games, the set of Nash equilibria is typically very large \citep{catalano2022network}. This induces substantial indeterminacy, and potentially major inefficiency, of economic outcomes. In relation to the macroeconomic analysis of production networks \cite[see e.g.][]{acemoglu2012network,baqaee2020productivity}, we show that, in the case of constant returns to scale, profit maximization can be equivalently expressed as the maximization of Domar weights. These weights defined as the ratio of a firm's output to GDP, underpin the standard aggregation result in growth accounting \cite[Hulten theorem, see][]{hulten1978} and have been widely adopted in models of shock propagation.


\subsection{Related literature} 

Our contribution builds on the recent thread of literature that analyses general equilibrium economies through the prism of network theory in order to generate new macroeconomic insights \citep{acemoglu2012network,gualdi2016emergence,moran2019may,carvalho2021supply,dessertaine2022out}. It is more specifically related to \citep{gualdi2019endogenous}
and \citep{acemoglu2020endogenous}, which subsume a standard general equilibrium model into a dynamic network formation process. Yet, these contributions focus on the interplay between network formation and economic growth rather than on the potential negative externalities induced by strategic behavior (see Subsection \ref{subsec:compare} below for an extensive comparison of our approach with that of \cite{acemoglu2020endogenous}). From a more technical perspective, our results echo the recent emphasis on Domar weights as a measure of macroeconomic relevance/centrality \cite[see e.g.][]{baqaee2020productivity}. Indeed, we find that the objective of firms, under constant returns to scale, can be expressed as the maximization of their Domar weight. Domar weights have proven to be central in understanding how sectoral productivity translates into aggregate outcomes via intermediate input linkage. For instance,  \citep{acemoglu2012network} demonstrate how large-Domar-weight sectors can serve as critical nodes in transmitting micro shocks to the macro-economy, while \citep{Carvalho2019} offer a comprehensive overview of how production network structures shape this amplification. Relatedly, a recent contribution by \cite{kopytov2024endogenous} investigates the formation of endogenous networks in the presence of supply chain risk and finds that more productive and stable firms have higher Domar weights. Yet, as \citep{acemoglu2020endogenous}, \cite{kopytov2024endogenous} consider cost-minimizing firms in a setting with constant returns to scale.  Comparatively, by considering profit maximization behavior and allowing for decreasing returns to scale, our approach offers new insights beyond the predictions of standard Domar-weight-based models.

Our approach can also be contrasted with that in the trade literature. In the canonical models of trade \cite[e.g.][]{eaton2002technology}, firms take as given costs, be they related to production or trade (e.g. iceberg costs) and are price-takers at equilibrium. In our setting, firms strategically anticipate the impact of their linkage choices on the general equilibrium structure and the prevailing prices. We thus offer a complementary perspective, beyond general equilibrium, where firms strategically choose their supply relationships rather than compare their relative costs.  Hence, beyond the formation of global supply chains, our framework can be used to model behavior that seeks to influence the structure of international trade flows, e.g. by lobbying on trade agreement or trade barriers (whose existence is well documented, see e.g. \cite{rodrik1995political,levy1999lobbying,stoyanov2009trade,bombardini2012competition}).

Our focus on the role of production networks in the propagation of risks is linked to a range of contributions about network-based amplification of micro-economic risks \citep{bak1993aggregate,battiston2007credit,barrot2016input,carvalho2021supply}. Yet, these contributions generally aim at quantifying the amplification of micro-economic shocks through network effects, whereas we are concerned with the micro-economic and behavioral determinants of the riskiness of the network. In this respect, our approach is closely related to \cite{elliott2022supply} that shows that endogenous formation of supply networks can be conductive to fragility. However, the approach of \cite{elliott2022supply} is less directly related to general equilibrium than ours, analyzes specifically tree-like supply chains, and focuses on fragility per se rather than on the interplay between strategic behavior and social efficiency.


Its emphasis on social efficiency relates our work to previous contributions that have investigated, in different contexts, negative externalities induced by network formation processes. 
\cite{fagiolo2005endogenous} investigates a model of network formation where externalities become negative as the size of an agent's neighborhood grows. \cite{carayol2008inefficiencies} analyze network formation in a spatialized version of the connections model of \citep{jackson1996strategic} and show that emergent networks are insufficiently dense and should be more structured around central agents. \cite{morrill2011network} considers a model where any new relationship imposes a negative externality on the rest of the network and shows that socially efficient and stable networks generally diverge.  \cite{buechel2012under} analyze systematically the sources of inefficiency in network formation and relate situations of positive externalities with stable networks that cannot be too dense, while situations with negative externalities tend to induce ``too dense''  networks. Accordingly, our results on efficient networks in the presence of exogenous shocks are qualitatively very similar to that of \cite{acemoglu2015systemic} in financial networks.   Indeed, we find that when risks are ``small", diversified, highly connected, network structures are more resilient to shocks. When risks are large and prone to amplification, the benefits of diversification can be offset by increasing risk exposure, and less connected network structures are more resilient.

\vspace{0.5cm}

The remainder of this paper is organized as follows. Section \ref{Section2} presents our model of production network formation and its connection to the general equilibrium theory. Section \ref{Section3} provides a network-based characterization of welfare and equilibrium behavior, investigates the relationship between equilibrium welfare and the distribution of productivity, analyzes the price of anarchy, and conducts a comparative analysis of the resilience of equilibrium production networks to disruptions caused by exogenous shocks. Section \ref{conclusion} concludes. An extended appendix contains the set of proofs for our results.

\color{black}
\section{Strategic formation of general equilibrium networks}\label{Section2}

\subsection{Economic framework}

We consider an economy with a representative household, indexed by $0,$ and a finite number of firms $M:=\{1,\cdots,m\} \subset \N^*$ producing differentiated goods. We let $N = M \cup \{0\}.$ The representative household supplies one unit of labor, receives the profits from the firms and is characterized by a Cobb-Douglas utility function of the form 
\begin{equation*} 
u(x_{0,1},\cdots,x_{0,n}):=\prod_{i\in M} x_{0,i}^{a_{0,i}},  
\end{equation*}
where $a_{0,i}$ denotes the share of firm $i$ in the household's consumption expenditure, and $\sum_{i \in M} a_{0,i} = 1$. Throughout this paper, we use bold notation to denote vectors; e.g. $\pmb{a}_0:= (a_{0,i})_{i\in M} \in \R^M_{+}$ denote the vector of consumption shares.\footnote{With innocuous abuse of notation, we use $\R^M$, $\R^{M\times M}$, $\R^N$ and $\R^{N\times N}$ denoting the space $\R^m$, $\R^{m\times m}$, $\R^{m+1}$, and $\R^{(m+1)\times (m+1)}$ respectively.}
 
As in \cite{acemoglu2020endogenous}, we assume that the firms have the ability to choose their suppliers and that these endogenous choices define their production technology. More specifically, we assume that the goods are grouped in a set $L$ of categories/sectors, and we denote by $M_{\ell}$ the set of firms producing goods of type $\ell$ (we consider $0$ to denote a specific category for labor and define $M_0=\{0\}$). Each firm $i$ is then characterized at a ``meta-technological" level by a vector of requirements $\pmb{b_i}:=(b_{i,0},\cdots,b_{i,L}) \in \R^L_+$  such that $\sum_{\ell \in L} b_{i,\ell} \leq 1$. In this setting, $b_{i,\ell}$ represents the expenditure share of firm $i$ to the input sector $\ell$. Firm $i$ must then choose its suppliers under these sectoral constraints, i.e. it must choose input shares $(a_{i,j})_{j \in M} \in \R_+^M$  towards each agent such that, for all $\ell \in L$,
\begin{equation*} 
\sum_{j \in M_{\ell}}  a_{i,j}= b_{i,\ell}. 
\end{equation*}
Its actual production technology is then given by a Cobb-Douglas production function of the form
\begin{equation*} 
f_{\pmb{a}_i}(x_{i,0},\cdots,x_{i,m}):=\lambda_i(\pmb{a}_{i}) \prod_{j\in N} x_{i,j}^{a_{i,j}},
\end{equation*}
where $\pmb{a}_{i}:= (a_{i,j})_{j\in N}$ denotes firm $i$'s choice of suppliers, and $\lambda_i(\pmb{a}_{i}) \in \R_{++}$ is a productivity parameter that depends on this choice.


The choice of each firm $i$ regarding its vector of suppliers, $\pmb{a}_i$, together with the household's consumption share vector $\pmb{a}_0$, defines a general equilibrium economy. Given that the utility and production functions are Cobb-Douglas, the economy is completely characterized by the productivity functions $\pmb{\lambda}:= (\lambda_i)_{i\in M}$, the consumption shares $\pmb{a}_0,$ and the collection of supplier vectors $\pmb{a}:= (\pmb{a}_i)_{i\in M}$. We shall thus denote this economy as $\mathcal{E}(\pmb{\lambda,a})$. A general equilibrium of the economy $\mathcal{E}(\pmb{\lambda,a})$ is standardly defined as follows:

\begin{defi} 
A general equilibrium of the economy $\mathcal{E}(\pmb{\lambda,a})$ is a collection of prices $\overline{\pmb{p}} \in \R^N_+$, (final and intermediary) consumption choices $\overline{\pmb{x}} \in \R_+^{N \times N}$ (with $\overline{x}_{0,0} = 0$), and output $\overline{\pmb{y}} \in \R^N_+$ (with $\y_0=1$) such that
\begin{enumerate}
\item each firm $i \in M$ maximizes its profit under the technological constraints, i.e. $(\overline{y}_i, \overline{x}_i)$ is solution of the following problem:
$$(\mathcal{P}): \left\{\begin{array}{cc} \max & \overline{p}_i y_i -\overline{\pmb{p}} \cdot \pmb{x}_i  \\ & \\ \text{s.t} & y_i = f_{\pmb{a}_i}(\pmb{x}_i)  \\ \end{array};\right.$$

\item the household maximizes its utility under its  budget constraint, i.e. $\overline{x}_0$ is solution of the following problem:
$$ (\mathcal{C}): \left\{\begin{array}{cc} \max & u(\pmb{x}_0) \\ & \\ \text{s.t} & \overline{\pmb{p}} \cdot \pmb{x}_0 \leq p_0+ \sum_{j \in M} \overline{p}_j \overline{y}_j -\overline{\pmb{p}} \cdot \overline{\pmb{x}}_j \\ \end{array};\right.$$

\item all markets clear, i.e. for all $i \in N,$ one has
$$ \sum_{j \in N}  \overline{x}_{j,i} = \overline{y}_i.$$
\end{enumerate}
\end{defi}

\subsection{Existence of a general equilibrium}
Let us recall that in a Cobb-Douglas setting, for all $i,j \in M,$ $a_{i,j}$ represents the proportion of firm $i$'s expenses directed towards firm $j.$ Furthermore, if $\sum_{k \in N} a_{i,k} \leq 1$, the profit rate of a profit-maximizing firm $i$ is given by the degree of decreasing returns to scale $\varepsilon_i =(1-\sum_{j \in N} a_{i,j} )= (1-\sum_{\ell =0}^L b_{i,\ell})$. Hence, if $p_0$ denotes the price of labor and $\pmb{v} := (v_i)_{i \in M}$ is the vector of firm revenues at equilibrium, then for all $j \in M$, it must hold that
 \begin{equation} 
 v_j= a_{0,j}p_0+ a_{0,j} \sum_{i \in M} v_i \varepsilon_i + \sum_{i \in M} a_{i,j} v_i, \label{balance-flow}
\end{equation}
where $a_{0,j}p_0$ corresponds to consumption of good $j$ based on labor income, $a_{0,j} \sum_{i \in M} v_i \varepsilon_i$ corresponds to consumption of good $j$ based on profit income, and $\sum_{i \in M} a_{i,j} v_i$ corresponds to intermediary consumption of good $j$ (see the proof of Proposition \ref{prop-equi} for details). Thus, Equation (\ref{balance-flow}) holds for all $j \in M$ is equivalent to that $\left(\begin{array}{c}p_0 \\\pmb{v} \end{array}\right)= \widetilde{A}^T \times  \left(\begin{array}{c}p_0 \\\pmb{v} \end{array}\right)$ where matrix $\tilde{A}$ is defined below and $X^T$ denotes the transpose of $X$.

%

\begin{equation*}
\widetilde{A}:= \left(\begin{array}{cccc}0 & a_{0,1} & \cdots &  a_{0,m} \\  a_{1,0}& a_{1,1} + \varepsilon_1 a_{0,1} & \cdots & a_{1,m} + \varepsilon_1 a_{0,m}   \\ \vdots & \vdots & \vdots & \vdots   \\ a_{i,0} & a_{i,1} + \varepsilon_i a_{0,1} & \cdots &   a_{i,m} + \varepsilon_i a_{0,m}\\  \vdots &  \vdots &  \vdots &  \vdots  \\  a_{m,0} & a_{m,1} + \varepsilon_m a_{0,1}& \cdots &   a_{m,m} + \varepsilon_m a_{0,m}  \end{array}\right).
\end{equation*}

The matrix $\widetilde{A}$, a natural extension of the standard input-output matrix $A:=(a_{i,j})_{M \times M}$, provides a comprehensive representation of the equilibrium financial flows in the economy. These flows typically fall into four categories: intermediary financial flows between firms $(a_{i,j})_{M \times M}$, payments for labor services $(a_{i,0})_{i \in M}$, consumption flows $(a_{0, i})_{i \in M}$, and financial flows based on profit $(\varepsilon_i a_{0,j})_{i \in M, j \in M}$. With decreasing return to scale, $a_{i,0}$ does not represent the complete financial flows between the firm and the household as it does not account for the profit distribution. Figure 1 below illustrates how the financial flows go from firm $i$ to firm $j$ at equilibrium, highlighting the extended structure captured by matrix $\widetilde{A}$.
\begin{figure}[h]
    \centering
    \resizebox{0.35\textwidth}{!}{%
\begin{tikzpicture}[node distance={25mm}, every node/.style={font=\small}] 
  \node[node in, draw] (0) {$0$}; 
  \node[node out] (1) [left of=0] {$i$}; 
  \node[node out] (2) [right of=0] {$j$}; 

  \draw[->, myred] (1) -- (0) node [midway,above] {$\varepsilon_i$}; 
  \draw[->, blue] (0) -- (2) node [midway,above] {$a_{0,j}$}; 
  \draw[->] (1) to[out=60, in=120, looseness=1] node[midway, below, yshift=-2pt] {$a_{i,j}$} (2);
\end{tikzpicture}}
\caption{\textit{\small\itshape  This figure shows the financial flows from firm $i$ to firm $j$. The direct flow (black arrow) represents firm $i$'s expenditure on intermediate goods $x_{i,j}$, indexed by $a_{i,j}$. The indirect flow arises from household consumption: the household buys the final good $x_{0,j}$ from firm $j$, contributing $a_{0,j}$ (\textcolor{blue}{blue arrow}). Firm $i$ supports this consumption by paying wages and profits to the household, contributing indirectly via its profit share $\varepsilon_i$ (\textcolor{myred}{red arrow}). This indirect contribution is thus $\varepsilon_i a_{0,j}$.}}
\end{figure}
\begin{remark} If there are constant returns to scale, one simply has $\widetilde{A}=\left(\begin{array}{cc}0 & a_0^T \\ a_{\cdot,0} & A\end{array}\right)$.
\end{remark}  

\noindent It is straightforward to check that $\widetilde{A}$ is row-stochastic. The following conditions then imply that it is aperiodic and irreducible.  
\begin{assump} \label{assump}The model parameters satisfy the following conditions:
\begin{enumerate}
\item The representative household consumes every good, i.e. $\pmb{a}_0 \in \R^M_{++}$. 
\item All firms use labor as input, i.e. $b_{i,0}=a_{i,0}>0$ for all $i \in M$. 
\item At least one firm uses another input than labor in its production process, i.e. there exists $i_0\in M$ and $\ell \in L/\{0\}$  such that $b_{i_0,\ell}>0$.
\end{enumerate}
\end{assump}

\begin{lemma} \label{ergodic} Under Assumption \ref{assump}, the matrix $\widetilde{A}$ is aperiodic and irreducible. 
\end{lemma}

\noindent This suffices to show the existence and to provide a characterization of the general equilibrium of $\mathcal{E}(\pmb{\lambda,a})$ 
\begin{prop} \label{prop-equi}
There exists a general equilibrium in the economy $\mathcal{E}(\pmb{\lambda,a})$ that is unique up to price normalization. Assuming $p_0=1,$ one further gets that
\begin{enumerate}
\item equilibrium revenues $\overline{v}_i:= \overline{p}_i \overline{y}_i$ are such that 
\begin{equation*}  
\left(\begin{array}{c}1 \\\overline{\pmb{v}} \end{array}\right)= \widetilde{A}^T \times  \left(\begin{array}{c}1 \\ \overline{\pmb{v}} \end{array}\right),
\end{equation*} which means that Equation (\ref{balance-flow}) holds and $\sum_{i \in M} a_{i,0} v_i=1$;
\item equilibrium profits are such that 
\begin{equation*}
\overline{\pi}_i=(1-\sum_{j\in N} a_{i,j}) \overline{v}_i= (1-\sum_{\ell \in L} b_{i,\ell}) \overline{v}_i;
\end{equation*}
\item equilibrium prices are such that 
\begin{equation}
\log(\overline{\pmb{p}})= (A-I)^{-1} \pmb{u} + (A-I)^{-1} D \log(\overline{\pmb{v}}) \label{eq-eqprice},
\end{equation}  
where $u_i= \log(\lambda_i)+  \sum_{j \in N} a_{i,j} \log(a_{i,j})$, $D=\diag (\sum_{j \in N} a_{i,j} -1)$, $\log(\p)_i=\log(\p_i)$, and $\log(\v)_i=\log(\v_i)$.
\end{enumerate}
\end{prop}

\subsection{Characterization of macro-economic aggregates}
\label{subsec:macro}

Building on the characterization of general equilibrium in Proposition \ref{prop-equi}, one can provide closed-form  analytical expressions for key macro-economic aggregates in the economy $\mathcal{E}(\pmb{\lambda,a})$.

\begin{enumerate}
\item First, nominal GDP is given by the sum of revenues of the household, denoted by $v_0$. Namely, one has
\begin{equation*}
v_0 =\p_0 + \sum_{i \in M}  \varepsilon_i \overline{v}_i,
\end{equation*} 
where $\overline{v}_i$ is defined in Proposition \ref{prop-equi}. We also consider the logarithm of nominal GDP in the economy, denoted by $\text{G}_n(\pmb{a},\pmb{\lambda})$, defined as:
\begin{equation*}
G_n(\pmb{a},\pmb{\lambda}) =\log \bigg( p_0 + \sum_{i \in M}  \varepsilon_i \overline{v}_i \bigg).
\end{equation*} 
 In the case of constant returns to scale, this further simplifies to 
\begin{equation*}
G_n(\pmb{a},\pmb{\lambda}) =\log(\p_0)
\end{equation*} 
as in \cite{acemoglu2012network}.

\item Using the geometric mean of prices as a price index \cite[as is standard practice in national accounts, see e.g.][]{dalton1998incorporating}, we obtain equilibrium real GDP. Namely, the log of real GDP in the economy $\mathcal{E}(\pmb{\lambda,a})$ is given by:

\begin{equation*}
G_r(\pmb{a},\pmb{\lambda}) = \log\bigg(\frac{\p_0 +  \sum_{i \in M}  \varepsilon_i \overline{v}_i}{\prod_{i \in M} \p_i^{\nicefrac{1}{m}}}\bigg) = \log\bigg(\p_0+  \sum_{i \in M}  \varepsilon_i \overline{v}_i\bigg)  -\frac{1}{m} \sum_{i=1}^m \log(\p_i).
\end{equation*}
Using Equation (\ref{eq-eqprice}), this can be equivalently written as: 
\begin{equation*}
G_r(\pmb{a},\pmb{\lambda}) = \log\bigg(\p_0+  \sum_{i \in M}  \varepsilon_i \overline{v}_i\bigg)  + \mathbf{1}  \dfrac{\alpha}{m}  [I-A]^{-1} \pmb{u}+ \mathbf{1}(I-A)^{-1} D \log(\overline{\pmb{v}}),
\end{equation*}
where $u_i= \log(\lambda_i)+  \sum_{j \in N} a_{i,j} \log(a_{i,j})$ and   $D=\diag (\sum_{j \in N} a_{i,j} -1)$. In the case of constant returns to scale and using the normalization $\p_0=1,$ this further simplifies to 
\begin{equation*}
G_r(\pmb{a},\pmb{\lambda}) = \mathbf{1}  \dfrac{\alpha}{m}  [I-A]^{-1} \pmb{u}.
\end{equation*}
That is the standard formula for real GDP used e.g. in \cite{acemoglu2012network}.

\item If prices are weighted proportionally to consumption shares, the geometric price index is given by $\prod_{i \in M} \p_i^{a_{0,i}},$ and one obtains an alternative formulation for real GDP:
\begin{equation*}
G'_r(\pmb{a},\pmb{\lambda}) = \log(\p_0)  + \pmb{a}_0^T   [I-A]^{-1} \pmb{u} + \pmb{a}_0^T (I-A)^{-1} D \log(\overline{\pmb{v}}).
\end{equation*}
In the case of constant returns to scale and using the normalization $\p_0=1,$ this further simplifies to 
\begin{equation*}
G'_r(\pmb{a},\pmb{\lambda}) = \pmb{a}_0^T   [I-A]^{-1} \pmb{u}.
\end{equation*}
\item Domar weights for sector/firm $i$ are a measure of sectoral importance  defined as the ratio between the firm's revenues and nominal GDP. In our setting, they are given for all $i \in M$ by:
\begin{equation*}
\beta_i=\dfrac{\v_i}{ \p_0 + \sum_{i \in M}  \varepsilon_i \overline{v}_i}.
\end{equation*}
Hence, in the case of constant returns to scale, Domar weights are identical to equilibrium revenues (up to the price normalization $\p_0=1$). In the general case, one can use Equation (\ref{balance-flow}) and get: 
\begin{align*}
\beta_i & =\dfrac{ a_{0,i}p_0+ a_{0,i} \sum_{j \in M} \v_j \varepsilon_j + \sum_{j \in M} a_{i,j} \v_j,}{\p_0 + \sum_{i \in M}  \varepsilon_i \overline{v}_i} \\ &=  \dfrac{a_{0,i} (\p_0 + \sum_{i \in M}  \varepsilon_i \overline{v}_i) + \sum_{j \in M} a_{i,j} \v_j}{\p_0 + \sum_{i \in M}  \varepsilon_i \overline{v}_i} \\ &=a_{0,i} + \sum_{j \in M} a_{i,j}\beta_j.
\end{align*} 
So that, the vector of Domar weights is obtained as the solution of the following matricial equation:
\begin{equation*}
 \pmb{\beta}=(I-A^T)^{-1}\pmb{a}_0.
 \end{equation*}

\item Finally, equilibrium utility is given by  $u(\x)=\displaystyle\prod_{i \in M} \left(\frac{a_{0,i} \overline{v}_0}{\overline{p}_i}\right)^{a_{0,i}}.$   Using the logarithm of  utility, $V(\pmb{a},\pmb{\lambda}),$ as a measure of social welfare in the economy, we get
\begin{equation*}
V(\pmb{a},\pmb{\lambda})=\log(\v_0)+\sum_{i \in M} a_{0,i}\log(a_{0,i})- (\pmb{a}_0)^T\log(\overline{\pmb{p}}).
\end{equation*}
Or equivalently, using Equation (\ref{eq-eqprice}),
\begin{equation*}
V(\pmb{a},\pmb{\lambda})=\log(v_0)+\sum_{i \in M} a_{0,i}\log(a_{0,i})+ (\pmb{a}_0)^T(I-A)^{-1} \pmb{u}  + (\pmb{a}_0)^T(I-A)^{-1} D \log(\overline{\pmb{v}}).
\end{equation*}
In the case of constant returns to scale, this further simplifies to
\begin{equation*}
V(\pmb{a},\pmb{\lambda})= \log(p_0)+\sum_{i \in M} a_{0,i}\log(a_{0,i})+ (\pmb{a}_0)^T(I-A)^{-1} \pmb{u} \end{equation*}
One can thus remark that, up to a constant, social welfare can be identified  with real GDP when the price index is weighted by consumption shares.    
\end{enumerate}

\subsection{Strategic framework}
The economic framework we consider is very similar to that of \cite{acemoglu2020endogenous}, where the choice of suppliers by firms also defines a general equilibrium economy. However, \cite{acemoglu2020endogenous} focus on the emergence of endogenous growth through increasing product variety in a setting with constant returns to scale, and where firms choose their suppliers in view of minimizing production costs. In this setting, the incentives of firms are aligned with those of the household, and firms thus somehow act as the agents of the household (see Subsection \ref{subsec:compare} for an extensive comparison between our approach and that of \cite{acemoglu2020endogenous}). We are rather concerned by the strategic behavior of profit-maximizing firms and the negative externalities they can induce on the household. 

As a benchmark representation of this situation, we consider the normal-form game $\mathcal{G}(\pmb{a}_0,\pmb{b})$ in which each firm $i \in M$

\begin{enumerate}
\item chooses as a strategy a vector of input weights $\pmb{a}_i$ in the set of admissible  technological configurations $S_i(\pmb{b}):= \{\pmb{a}_i \in \R^N_+ \mid \sum_{j \in M_{\ell}}  a_{i,j}= b_{i,\ell} \}$ and
\item receives as payoff the equilibrium profit in the hence defined economy $\mathcal{E}(\pmb{\lambda,a})$, i.e.
\begin{equation*} 
\pi_i(\pmb{a}_i,\pmb{a}_{-i})= (1-\sum_{j\in N}  a_{i,j}) \v_i =  (1-\sum_{ \ell \in L} b_{i,\ell}) \v_i = \varepsilon_i \v_i,  
\label{def-prof}
\end{equation*}
where $\overline{\pmb{v}}$ is such that 
\begin{equation*}  
\left(\begin{array}{c}1 \\\overline{\pmb{v}} \end{array}\right)= \widetilde{A}^T \times  \left(\begin{array}{c}1 \\ \overline{\pmb{v}} \end{array}\right),
\end{equation*} 
and $\sum_{j \in M} a_{j,0} \overline{v}_j=1$ by Proposition \ref{prop-equi}.
\end{enumerate}

\noindent A Nash equilibrium in the game $\mathcal{G}(\pmb{a}_0,\pmb{b})$ is defined as follows:

\begin{defi} A Nash equilibrium of the game $\mathcal{G}(\pmb{a}_0,\pmb{b})$ consists in the choice by each firm $i \in M$  of a vector of input weights $\overline{\pmb{a}}_i$ in the set of admissible technological configurations $S_i(\pmb{b})$ such that for all $i \in  \M$ and all $\pmb{a}_i \in S_i(\pmb{b}),$ one has
\begin{equation*}  
\pi_i(\overline{\pmb{a}}_i,\overline{\pmb{a}}_{-i})\geq \pi_i(\pmb{a}_i,\overline{\pmb{a}}_{-i}).
\end{equation*}
\end{defi}

Hence, we represent the formation of production networks as a network formation game in which each firm chooses its suppliers with the aim of maximizing profit at the corresponding economic equilibrium. In economic terms, firms choose their position in the global supply chain, assuming that supply relationships will remain stable and that competition will occur through (partial) substitution of commodities. The relationships between economic outcomes and network characteristics prevailing in the economy $\E(\pmb{a},\pmb{\lambda})$ imply alternative characterizations of the strategic objective of the firms and of their equilibrium behavior. First, given equilibrium profits are proportional to equilibrium revenues, the objective of the firm can be equivalently defined as that of maximizing its revenues. This notably extends the definition of the game to the limiting case of constant returns to scale. In that limiting case, as noted in $\text{(iv)}$ of Subsection \ref{subsec:macro}, equilibrium revenues are equal to Domar weights. Hence, the objective of the firm can equivalently be expressed, in the case of constant returns to scale, as the maximization of its  Domar weight. From a network perspective, equilibrium revenues are given by the eigenvector centrality of firms $\v$ for the matrix $\widetilde{A}^T.$ Hence, $\mathcal{G}(\pmb{a}_0, \pmb{b})$ is equivalent to the network formation game where each player $i \in M$ chooses its outgoing links in $S_i(\pmb{b})$ so as to maximize its eigenvector centrality. In turn, as put forward in \cite{catalano2022network}, this is equivalent to the game where player $i$ maximizes its invariant probability under the Markov chain with transition matrix $\widetilde{A}^T$. These various characterizations appear as instances of a core objective of the firm that is to maximize its relative importance in the economy.

\subsection{Equilibrium behavior}\label{Net_based_equi}

In this subsection, we exploit the fact that $\mathcal{G}(\pmb{a}_0, \pmb{b})$ is equivalent to a game where firms choose their outgoing network connections to maximize their eigenvector centrality in order to characterize precisely the micro-economic behavior of firms and to show the existence of a Nash equilibrium.  
In our setting, the eigenvector centrality of firms can be characterized in terms of network walks (or equivalently financial flows) emanating from the household. Namely, let us associate to a link $(i,j)$ the weight $\tilde{a}_{j,i} := a_{i,j} + \varepsilon_i a_{0,j}$ corresponding to the share of revenues of $j$ that comes from the firm $i$, either directly through intermediary consumption $a_{i, j}$ or through the consumer via the spending of profit-related income $\varepsilon_i a_{0, j}$ and denote the  associated matrix as $\widetilde{A}_M$.\footnote{Remark that $\widetilde{A}_M^T$ is the matrix obtained from matrix $\tilde{A}$ by deleting the first row and the first column.}

\begin{figure}[h]
    \centering
    \resizebox{0.4\textwidth}{!}{%
\begin{tikzpicture}[node distance={25mm}, every node/.style={font=\small}] 
  \node[node in, draw] (0) {$0$}; 
  \node[node out] (1) [left of=0] {$i$}; 
  \node[node out] (2) [right of=0] {$j$}; 

  \draw[->, myred] (1) -- (0) node [midway,above] {$\varepsilon_i$}; 
  \draw[->, blue] (0) -- (2) node [midway,above] {$a_{0,j}$}; 
  \draw[->] (1) to[out=60, in=120, looseness=1] node[midway, below, yshift=-2pt] {$a_{i,j}$} (2);
\end{tikzpicture}}
\caption{\textit{\small\itshape  This figure shows the financial flows from firm $i$ to firm $j$. The direct flow (black arrow) represents firm $i$'s expenditure on intermediate goods $x_{i,j}$, indexed by $a_{i,j}$. The indirect flow arises from household consumption: the household buys the final good $x_{0,j}$ from firm $j$, contributing $a_{0,j}$ (\textcolor{blue}{blue arrows}). Firm $i$ supports this consumption by paying wages and profits to the household, contributing indirectly via its profit share $\varepsilon_i$. This indirect contribution is thus $\varepsilon_i a_{0,j}$.}}
\end{figure}

Accordingly, given a walk of length $k$ from $i \in M$ to $j \in M$ denoted as $p = (h_1, h_2, \dots, h_k) \in M^k$ with $h_1 = i$ and $h_k = j$, let us define the associated weight as $w_{\pmb{a}}(p):= \prod_{i=1}^{k-1} (a_{h_i, h_{i+1}} + \varepsilon_{h_i} a_{0, h_{i+1}}) = \prod_{i=1}^{k-1} \tilde{a}_{h_{i+1}, h_i}$, representing the share of revenues of $i$ reaching $j$ through the corresponding walk. We then denote by $\mathcal{P}_{j,i}$ the set of walks from $j \in M$ to $i \in M$ and by $P_{j,i}(\pmb{a})=  \sum_{p \in \mathcal{P}_{j,i}} w_{\pmb{a}}(p)$ the sum of weights of walks in $\mathcal{P}_{j,i}.$ Now, one can then express the profit of firms in terms of the weights of network walks.

\begin{lemma} \label{prof2} For all $i \in M$, 
$\pi_i(\pmb{a})=\varepsilon_i (I-\widetilde{A}_M)^{-1}_i \pmb{a}_0=  \varepsilon_i \sum_{j \in M} a_{0,j} P_{j,i} (\pmb{a}).$
\end{lemma}

\begin{remark}
Being row-stochastic, $\widetilde{A}^T$ is the transition matrix of a Markov chain, and $P_{j,i}$ can be interpreted in the context of this Markov chain as the probability of reaching $j$ from $i$ before reaching $0$ (considering only walks with nodes in $M$).
\label{markovrem}
\end{remark}

One can further decompose network walks  into direct walks and cycles. A walk $p = (h_1,h_2,\dots ,h_k)$ is called a direct walk from $j$ to $i$ if  $h_j \neq i$ for all $j \in \{2,\dots,k-1 \}$ and a direct walk from $i$ to $i$ is called a direct cycle. We use $\mathcal{D}_{j,i}$ to denote the set of all direct walks from $j$ to $i,$ (in particular, $\mathcal{D}_{i,i}$ is the set of direct cycles around $i$) and $D_{j,i}(\pmb{a})$ to denote the sum of weights of such walks. One shall note that $D_{j,i}(\pmb{a})$ is independent of $a_i$ as it contains no outgoing link from $i$. The profit of firm $i$ can be expressed in terms of the direct paths to $i$ and the direct cycles around $i$ as follows:
 
\begin{lemma} \label{profchar} The profit function of firm $i \in M$ is given by
\begin{align*}
\pi_i(\pmb{a})= \varepsilon_i P_{0,i}(\pmb{a}) =\varepsilon_i \dfrac{a_{0,i} +\sum_{j \in M \setminus\{i\}} a_{0,j} D_{j,i}(\pmb{a})}{1-D_{i,i}(\pmb{a})},
\end{align*}
or equivalently,
\begin{align*}
\pi_i(\pmb{a}_i,\pmb{a}_{-i})=\varepsilon_i \dfrac{a_{0,i} +\sum_{j \in M \setminus\{i\}} a_{0,j} D_{j,i}(\pmb{a}_{-i})}{1-\tilde{a}_{i,i}-\sum_{j \in M \setminus\{i\}} \tilde{a}_{i,j} D_{j,i}(\pmb{a}_{-i})}.
\end{align*} 
Moreover, $\pi_i$ is continuous and quasi-concave in $\pmb{a}_i.$ 
\end{lemma}

Overall, the revenues of firm $i$ correspond to the flow of consumption spending  that reaches $i.$ The profit of firm $i$ corresponds to a share $\varepsilon_i$ of these revenues. Lemmas \ref{prof2} and \ref{profchar} hence highlight that the firms aim to maximize incoming connectivity from the consumer. As the firms only choose their suppliers, they can only affect  this connectivity indirectly by maximizing the term  $D_{i,i}(\pmb{a}),$ i.e. by ensuring that the largest possible share of incoming value from the consumer remains in their supply chain (cf. the first expression of profit in Lemma \ref{profchar}). In other words, all firms seek to minimize financial outflows towards supply chains that are directed (upstream) towards other firms. Accordingly, the proof of Proposition \ref{prop-nash} underlines that if one considers only the acyclic part of supply chains, all firms have the common objective to minimize total financial flows so as to maximize their relative importance, which they can not directly influence strategically. Accordingly the sum of financial flows in the acyclic part of supply chains can be seen as a potential for the game. More precisely,  the existence of a Nash equilibrium in $\mathcal{G}(\pmb{a}_0, \pmb{b})$  follows from the observation that the game is an \textit{ordinal potential} (see \citep{monderer1996potential}), as in the closely related network formation games considered in \citep{catalano2022network} and \citep{cominetti2022buck}. 

\begin{prop}  The game $\mathcal{G}(\pmb{a}_0,\pmb{b})$ is ordinal potential and admits at least one Nash equilibrium. 
\label{prop-nash}
\end{prop}

Moreover, from the second expression of profit in Lemma \ref{profchar}, when considering suppliers in sector $\ell$, firm $i$ will select among those suppliers $j \in M_{\ell}$ for which $D_{j,i}(\pmb{a}) = \max_{k \in M_{\ell}} D_{k,i}(\pmb{a})$, which are the firms that have the highest share of revenues ``directly" reaching $i$.

From a more systemic perspective, in order to limit the outflow of money, firms have incentives to form clusters of suppliers with integrated supply chains. In particular, firms will always choose to keep production integrated internally if this is an available option, i.e. a firm does not have an external supplier for its own category of product.
\begin{prop} 
For every Nash equilibrium $\overline{\pmb{a}}$ of $\mathcal{G}(\pmb{a}_0,\pmb{b}),$ one has for all $i \in M$, 
\begin{equation*} 
i \in M_\ell \Rightarrow \a_{i,i} = b_{i,\ell}.
\end{equation*}\label{ownsupply} 
\end{prop}

\subsection{Relation to existing models of production network formation}
 \label{subsec:compare}
 
As emphasized above, the framework we consider is very similar to that of \cite{acemoglu2020endogenous} in the sense that we represent the choices of firms at a ``meta"-technological level that then determine the production structure of a general equilibrium economy. However, we explore very different behavioral determinants for these meta-technological choices and the resulting network formation  process. In \cite{acemoglu2020endogenous}, the firms' objective is to minimize production costs while in our framework firms are profit maximizers. These behaviors may be equivalent in a standard general equilibrium framework, but they are not necessarily so in our setting, where the realm of choices extends beyond that framework. Indeed, in the first decision step, when their interactions define the structure of the general equilibrium economy, firms might have  incentives to link to less productive/more expensive suppliers to latter exploit limited substitutability, pass the costs to the consumer and increase their revenues.In order to highlight the impact of  these strategic incentives, let us consider the impact of alternative assumptions on elasticities of substitution.\footnote{See Appendix B for a basic example illustrating the situation.} If intermediary inputs were perfect substitutes (and the corresponding part of the production function was linear), only the cheapest firms would sell intermediates at equilibrium and thus firms would have incentives to target minimal production costs during the network formation step in order to increase their revenues in the resulting general equilibrium. Oppositely, if intermediary inputs were perfect complements (and the corresponding part of the production function was Leontieff), the value of intermediate sales at equilibrium  would increase with prices and thus firms would have incentives to target maximal production costs during the network formation step in order to increase their revenues in the resulting general equilibrium. In the Cobb-Douglas case we consider, firms are ``neutral" about production costs: as highlighted in Proposition \ref{prop-equi}, the profits in the general equilibrium economy $\mathcal{E}(\pmb{\lambda}, \pmb{a})$ and, consequently, the payoffs in the game $\mathcal{G}(\pmb{a}_0, \pmb{b})$ are independent of the productivity functions $\pmb{\lambda}$. This implies in particular that firms do not have incentives to minimize production costs as opposed to \cite{acemoglu2020endogenous}. This contrast is further emphasized by the fact that in \cite{acemoglu2020endogenous}, a first welfare theorem holds (Theorem $3$) whereas in our setting there can be substantial negative externalities from strategic behavior, and inefficiencies at equilibrium, as illustrated in Subsection \ref{Welfare_prod}. From a more formal perspective, our approach differs from that of \cite{acemoglu2020endogenous} as we consider Nash equilibria where firms anticipate the impact of their actions on the general equilibrium that will prevail whereas \cite{acemoglu2020endogenous} considers more myopic firms that take the prevailing price as given as they do not anticipate the impact thereupon on their choice of suppliers.

\section{Social welfare in strategic production networks}\label{Section3}
In this section, we explore the potential inefficiencies induced in our framework by the discrepancy between micro-level incentives and social objectives. Considering potential applications to international trade within the context of the globalization/deglobalization debate \citep{witt2019globalization, kornprobst2021globalization}, we concentrate our analysis on a class of replicate economies. In this setting, where the determinants of final demand and social welfare are given by the representative utility function, we investigate how strategic choices of firms shape trade in intermediaries and global supply chains. We further investigate the welfare impacts of these strategic choices in the presence of risk and heterogeneous productivity. In this section, we mainly focus on the case of constant returns to scale.  

\subsection{Network-based characterization of the social welfare} 
In this subsection, we characterize social welfare in terms of network characteristics and highlight potential drivers of social inefficiency. 
As emphasized in Subsection \ref{subsec:macro}, social welfare is given by  $$V(\pmb{a},\pmb{\lambda})= \log(u(\x))=\log(v_0)+\sum_{i \in M} a_{0,i}\log(a_{0,i})+ (\pmb{a}_0)^T(I-A)^{-1} \pmb{u}  + (\pmb{a}_0)^T(I-A)^{-1} D \log(\overline{\pmb{v}}).$$
In the case of constant returns to scale, or as an approximation for small  $\varepsilon,$ this can be simplified into:
$$V(\pmb{a},\pmb{\lambda})= \log(p_0)+\sum_{i \in M} a_{0,i}\log(a_{0,i})+ (\pmb{a}_0)^T(I-A)^{-1} \pmb{u},$$
where for all $ i \in M,$  $u_i= \log(\lambda_i)+  \sum_{j \in N} a_{i,j} \log(a_{i,j}).$
\medskip

\noindent Normalizing $p_0$ to $1$ and noting that $\sum_{i \in M} a_{0,i}\log(a_{0,i})$ is fixed, we can analyze social welfare through the function\begin{equation} W(A,\pmb{\lambda}):=(\pmb{a}_0)^T(I-A)^{-1} \pmb{u}. \label{eq-welfsimple}\end{equation}
%

Equation (\ref{eq-welfsimple}) highlights two determinants of social welfare. First, the vector $u$ whose coordinates $u_i= \log(\lambda_i)+  \sum_{j \in N} a_{i,j} \log(a_{i,j}) $ correspond to the difference between the log productivity of firm $i$ and the Shannon entropy of its distribution of input weights across suppliers. The derivation of Equation (\ref{eq-prod2}) in the proof of Proposition  \ref{prop-equi} highlights how this ``entropy corrected productivity" emerges as a condensed measure of the production costs of firm $i$ combining productivity per se (measured by $\lambda_i$) and the complexity of the production process (measured by the Shannon entropy $ \sum_{j \in N} a_{i,j} \log(a_{i,j}) $). The larger the $u_i,$ the lower the production cost and  the equilibrium price of good $i$ shall be (all other things being equal, see the proof of Proposition  \ref{prop-equi}). Second, social welfare depends on the connectivity between the consumer and  firms with high ``entropy corrected productivity". Indeed, Equation (\ref{eq-welfsimple}) can be written as a weighted sum of walks in the production networks
\begin{equation*} 
W(A,\pmb{\lambda}):=  \sum_{i\in M}  \sum_{j\in M} a_{0,i} P_{i,j}(\pmb{a}) u_j.
\end{equation*}

\noindent Accordingly, the first-order conditions for welfare maximization can be expressed in terms of network structure as follows.
\begin{prop}  
If the production network $A$ is maximizing the welfare measure $W(A,\pmb{\lambda})$ among the set of  technological configurations 
$\{ A \in \R^{M \times M}_+ \mid \forall i \in M, \ \sum_{j \in M_{\ell}}  a_{i,j}= b_{i,\ell} \},$ one has for all $i \in M,$ for all $ \ell \in L$, and for all $j \in M_{\ell}$,

\begin{equation*} 
a_{i,j} >0 \Rightarrow \dfrac{\partial W(A,\pmb{\lambda})}{ \partial a_{i,j}} (\pmb{a})=P_{0,i}\left( \sum_{k \in M} P_{j,k}u_k+\log(a_{i,j})+1\right) = \max_{k \in M_\ell} \dfrac{\partial U}{ \partial a_{i,k}} (\pmb{a}).
\end{equation*}\label{prop-firstorderwelfare}
\end{prop}

Hence, the marginal contribution of link $(i,j)$ to social welfare depends on the connectivity between the household and firm $i$ on one hand, and the connectivity between firm $j$ and other firms weighted proportionally to (entropy-corrected) productivity, on the other hand. Accordingly, at a social optimum, only production links that maximize the connectivity between the household and high-productivity firms shall be enabled. 
\medskip
 
Overall, one observes major differences between the determinants of the social welfare and these of firms' profitability. The social welfare depends on connectivity from high (entropy-corrected) productivity firms towards the household whereas firms' profitability is independent of productivity and depends on the connectivity from the household towards the firms. In other words, the analysis of efficiency in the game $\mathcal{G}(\pmb{a}_0,\pmb{b})$ shall be based on a measure of social welfare defined independently from the individual payoff functions. This discrepancy between individual and social objectives can be observed in a number of network formation games, e.g. when individuals choose their peers in social networks to maximize influence but hence favor polarization \cite[see e.g.][]{bolletta2020polarization} or when financial institutions form lending relationships to diversify risk but thereby create structures prone to systemic risk \cite[see e.g.][]{allen2000financial}.

\subsection{Equilibria in a replicate economy}

In this subsection, we investigate in details the structure of the set of equilibria in the setting of a replicate economy \cite[see e.g.][]{debreu1963limit}. Every replication can be interpreted e.g. as a  country in a context of international trade. 

Let us first define a simple game as a game in which $M=L$ so that there exists exactly one firm in each category. Such a game is characterized by a share of consumption expenditures $\pmb{a}_0 \in \R^L_+$ and technological requirements $\pmb{b}_\ell \in \R^{L}$ for each firm $\ell \in L=M$. In the corresponding game $\mathcal{G}(\pmb{a}_0,\pmb{b})$, the strategy sets are singletons and thus the unique Nash equilibrium is characterized by $\overline{\pmb{a}}_\ell=\pmb{b}_\ell$ for all $\ell \in M$. 

We then define $\mathcal{G}^n(\pmb{a}_0, \pmb{b})$ as the $n$-fold replicate of $\mathcal{G}(\pmb{a}_0, \pmb{b}),$ where there are exactly $n$ firms in each category $\ell$, all with technology requirements $\pmb{b}_{\ell}$, and the household spending share on each firm in category $\ell$ is exactly $\nicefrac{a_{0,\ell}}{n}$. More formally,
\begin{defi} 
The $n$-fold replicate of the game $\mathcal{G}(\pmb{a}_0,\pmb{b}),$ denoted by $\mathcal{G}^n(\pmb{a}_0,\pmb{b}),$ is the game  $\mathcal{G}(\pmb{a}^n_0,\pmb{b}^n)$ such that 
\begin{enumerate}
 \item  the set of firms is $M=\{1, \dots, nL\}$;
 \item  the set of firms in category $\ell \in L$ is $M_\ell=\{(\ell-1)n+1, \dots, \ell n\}$;
 \item the consumption  shares $\pmb{a}_0^n$ are such that for all $\ell \in L$ and all  $i \in M_\ell,$ $a^n_{0,i}=\frac{a_{0,\ell}}{n}$;
 \item the technology requirement $\pmb{b}^n$ are such that  for all $i \in M_\ell,$ $b^n_i=b_\ell$.  
\end{enumerate}
\end{defi}

Such a replicate economy can be considered as the model of an economy with $n$ countries, where firms of country $c$ are labeled as $c, n+c, \dots, (L-1)n+c$. Each firm must choose in which country it sources its different inputs. Two polar cases can naturally be considered in this setting: one of an autarkic/island economy where each firm only sources inputs domestically, and one of a globalized economy where each firm sources inputs uniformly from each country. More broadly, one can consider partitions of the set of countries into economically integrated regions where firms source inputs only within the region. Let $\mathcal{Q}:=\{Q_1,\dots,Q_K\}$ be a partition of $\{1,\dots,n\}$. We define $Q_{\ell,k} = \{(\ell-1)n + i: i \in Q_k\}$ the set of replications of input type $\ell$ in the cluster $Q_k$. In an $n$-fold replicate game $\mathcal{G}^n(\pmb{a}_0,\pmb{b})$, we define a $\mathcal{Q}$-clustered network/production structure $\pmb{a}^\mathcal{Q}$ such that for all $i \in Q_{\ell,k}$,
\begin{equation}
\begin{cases}
a^\mathcal{Q}_{i,i}=b_{i,\ell},\\
a^\mathcal{Q}_{i,j}=0 \text{ for }  j \in Q_{\ell,k} \setminus \{i\}, \\
a^\mathcal{Q}_{i,j}=\dfrac{b_{i,\ell'}}{| Q_k|} \text{ for } j \in Q_{\ell',k} \text{ with } \ell' \neq \ell,\\
a^\mathcal{Q}_{i,j}=0 \text{ otherwise.} 
\end{cases} \label{clusters}
\end{equation}

It can be demonstrated that every $\mathcal{Q}$-clustered network/production structure is a Nash equilibrium of the $n$-fold replicate game $\mathcal{G}^n(a_0,b)$. Before stating our result, note that the $\mathcal{Q}$-clustered networks exhibit several key empirical features observed in real-world production networks. First, our equilibrium networks are consistent with the empirical observation documented by \cite{Alonso2024} that firms often have multiple trading partners from the same sector. Our $\mathcal{Q}$-clustered network enables multiple connections within each cluster, capturing the realistic possibility that firms link to several partners within a narrow production domain. Second, the model embeds the locality of connections, as firms are more likely to match within the same cluster, which reflects sectoral or geographic proximity, as emphasized in \cite{bernard2019production}. Moreover, the model generates a positive correlation between in- and out-degrees, a stylized fact observed by \cite{Lafondetat2023}, since firms with many suppliers also tend to be attractive as suppliers.

\begin{prop}\label{K-cluster} 
If, for all $i \in M$, $\varepsilon_i$ is sufficiently small, then every $\mathcal{Q}$-clustered network/production structure $\pmb{a}^\mathcal{Q}$ is a Nash equilibrium of the normal-form game $\mathcal{G}^n(\pmb{a}_0,\pmb{b})$.
\end{prop}

\begin{remark}
It has been previously observed that network formation games based on centrality can generate  a large number of Nash equilibria \cite[see e.g.][]{catalano2022network,cominetti2022buck}.
\end{remark}

Despite the built-in symmetry in the construction of  Q-clustered networks, they exhibit some key empirical features observed in real-world production networks. First, they are consistent with the empirical observation documented by \cite{alfaro-urena_zacchia_2024} that firms often have multiple trading partners from the same sector. Our Q-clustered framework enables multiple connections within each cluster, capturing the realistic possibility that firms link to several partners within a narrow production domain. Second, the model embeds the locality of connections, as firms are more likely to match within the same cluster---reflecting sectoral or geographic proximity, as emphasized in \cite{bernard2019production}. Moreover, our model generates a positive correlation between in- and out-degrees, a stylized fact observed by \cite{Lafondetat2023}, since firms with many suppliers also tend to be attractive as suppliers.

\subsection{Social welfare and productivity}\label{Welfare_prod}

Proposition \ref{K-cluster} implies that there is substantial indeterminacy about the network structure that can emerge from strategic interactions in the formation of global supply chains. Certain equilibrium configurations might lead to substantial inefficiencies. In order to investigate this issue, we shall quantify, in the following, these inefficiencies using the notion of price of anarchy. In our setting, it is defined as follows:

\begin{defi} The price of anarchy $POA(\pmb{b})$ is defined as the ratio between the largest equilibrium social welfare in the class of economies $\mathcal{E}(\pmb{a},\pmb{\lambda})$ with technology configurations in $S(\pmb{b})$ and the lowest  equilibrium social welfare in an economy that is a Nash equilibrium of the game $\mathcal{G}(\pmb{a}_0,\pmb{b}).$ 
\end{defi}

Among the potential equilibrium network configurations presented in Proposition \ref{K-cluster}, two polar cases will be of particular interest: (i) the \textit{islands economy} such that $|\mathcal{Q}|=n$, in which firms are only connected to firms in the same cluster/country, and (ii) the \textit{fully connected economy} such that $|\mathcal{Q}|=1$, in which firms are uniformly connected across clusters.

In the simple setting of the game $\mathcal{G}^n(\pmb{a}_0,\pmb{b})$, the impact of network structure on social welfare is mostly determined through the productivity characteristics of technologies, which are captured by the productivity functions $\lambda$. A benchmark case is given by Hicks-neutral productivity $\overline{\lambda}$ such that for all $\pmb{a}_i \in \R^M_+$,

\begin{equation*} 
\overline{\lambda}(\pmb{a}_i)=\frac{1}{\prod_{j\in M}a_{i,j}^{a_{i,j}}}.
\end{equation*}

Indeed, Hicks-neutral productivity exactly compensate productivity losses from increasing process complexity through positive spillovers on productivity. If all firms have Hicks-neutral productivity, social welfare is independent of network structure.

\begin{prop}\label{prop_hick_neutral} 
If all firms have Hicks-neutral productivity, i.e. for all $i \in M$, $\lambda_i=\overline{\lambda}$, the the social welfare is equal at all the $\mathcal{Q}$-clustered network equilibria of the normal form game $\mathcal{G}^n(\pmb{a}_0,\pmb{b})$. 
\end{prop}

The comparative statics between equilibria in the general case then depend on the relative strengths of technological spillovers on productivity with respect to the Hicks-neutral case. Namely, we say that a firm has increasing (resp. decreasing) \textit{returns to diversification} if positive spillovers from input diversification on productivity are greater (resp. smaller) than in the Hicks-neutral case. 

\begin{defi} Firm $i$ has increasing returns to diversification if for all $\pmb{a}_i,\pmb{a}'_i \in \R^M_+$,
$$ \overline{\lambda}(\pmb{a}_i)  \geq \overline{\lambda}(\pmb{a}'_i) \Rightarrow  \dfrac{\lambda_i(\pmb{a}_i)}{\lambda_i(\pmb{a}'_i)} \geq \dfrac{\overline{\lambda}(\pmb{a}_i)}{\overline{\lambda}(\pmb{a}'_i)}.$$
Respectively, firm $i$ has decreasing returns to diversification  for all $\pmb{a}_i,\pmb{a}'_i \in \R^M_+$,
$$ \overline{\lambda}(\pmb{a}_i)  \geq \overline{\lambda}(\pmb{a}'_i) \Rightarrow  \dfrac{\lambda_i(\pmb{a}_i)}{\lambda_i(\pmb{a}'_i)} \leq \dfrac{\overline{\lambda}(\pmb{a}_i)}{\overline{\lambda}(\pmb{a}'_i)}.$$
\label{return-diversification}
\end{defi}

\begin{remark} It is, in particular, the case that a firm has decreasing returns to diversification if its productivity is constant and independent of the network structure. 
\end{remark}

In a setting where there are increasing returns to diversification, more interconnected network structures will be more socially efficient, and conversely in the case where there are decreasing returns to diversification. In particular, one has an exact comparative static result comparing the islands and the fully connected economy:

\begin{prop}\label{prop_diversification} In the game $\mathcal{G}^n(\pmb{a}_0,\pmb{b})$,
\begin{itemize}
\item if all the firms have increasing returns to diversification, social welfare in the fully connected economy is greater than social welfare in the islands economy; 
\item if all the firms have decreasing returns to diversification, social welfare in the islands economy is greater than social welfare in the fully connected economy. 
\end{itemize}
\end{prop}   

These differences in welfare can be quite substantial from a quantitative standpoint, as highlighted by the following example:

\begin{prop} \label{poa-example} 
In the game $\mathcal{G}^n(\pmb{a}_0,\pmb{b})$ where all firms have constant productivity $\lambda,$ the welfare difference between the island and the fully connected economy is of the form $$K \log(n)$$ with $K>0.$ Accordingly, the price of anarchy tends to infinity as $n$ tends towards infinity.
\end{prop}

Proposition \ref{poa-example} highlights that the pure strategic focus of firms in our framework can have substantial consequences in terms of welfare. Indeed, except in the case of Hicks neutral productivity, there will be substantial differences in terms of welfare between the islands and the fully connected economy. The exclusive focus of  the firm on its profitability imply that both situations can always be sustained as Nash equilibria. This multiplicity of equilibria  with very different outcomes in terms of welfare  imply that the price of anarchy can be very large in our setting.  In this respect, let us emphasize that the lower bound on the price of anarchy provided in Proposition \ref{poa-example} is tight as social optimum could be even greater than the welfare prevailing at the best equilibrium.    

Overall, the results in this subsection emphasize that the optimal network structure depends on productivity spillovers. If there are strong productivity spillovers to input diversification, then more interconnected, globalized  network structures are more socially efficient. If these spillovers are weak, more localized networks structures are more efficient. Note however that our setting \`a la Armington is likely to underestimate the benefits of globalized networks because its  structure limits the possibilities of substitution between domestic and global goods.
 
Our results also complement those in \cite{acemoglu2020endogenous} giving necessary conditions for sustained economic growth via endogenous formation of production networks. Indeed, if there are increasing returns to diversification, total output and social welfare at the fully connected equilibrium  of  $\mathcal{G}^n(\pmb{a}_0,\pmb{b})$ will grow as $n$ grows. Hence, as in \cite{acemoglu2020endogenous}, the ability of firms to update their supply relationships following the entry of new firms, suffices to sustain economic growth. There is the choice between multiple suppliers, only those such that this is maximal shall be selected (or all suppliers shall have same marginal contribution towards household).

\subsection{Social welfare and risk}

The recent literature has emphasized (emerging) risks as potential major drivers of change in the structure of global supply chains \cite[see e.g.][]{razin2020globalization,irwin2020pandemic}. In order to investigate this issue, we extend our model by considering 
that each link $(i,j)$ can be, independently, disrupted with a probability $r_{i,j} \in [0,1]$. These exogenous shocks on supply relationships impact are assumed to impact the production process and thus the output of the firms. We formalize these impacts by considering that  if the set of links $K$  is disrupted in the network $A$, the production function of firm $i$ is altered to
$$ f_{\pmb{a}_i}(x_{i,0},\cdots,x_{i,n}):= (1-\rho)^{\phi_i(K,A)} \lambda_i(\pmb{a}_i) \prod_{j\in N} x_{i,j}^{a_{i,j}},$$
where $\rho$ is a parameter determining the average magnitude of shocks and $\phi$ is a disruption function defining the impact on output through aggregation of the shocks occurring in the supply chain. 
\medskip

One can remark that these disruptions of productivity do not impact the profits of firm at general equilibrium and thus do not modify the set of Nash equilibria of the game $\mathcal{G}^n(\pmb{a}_0,\pmb{b}).$ However, these shocks induce substantial modifications of social welfare, which is now given by the expected utility of the representative household given the distribution of risk, i.e.

\begin{equation}  \label{risk-welfare1} 
\widehat{V}(\pmb{a},\pmb{\lambda}) = \sum_{ K  \subset M \times M } [\prod_{\{ (i,j) \in K^c \}} (1-r_{i,j}) \prod_{\{ (h,k) \in K \}} r_{h,k}]  V(\pmb{a}, (1-\rho)^{\phi(K,A)} \otimes \lambda),
\end{equation}
and
\begin{equation} \label{risk-welfare2} 
\widehat{W}(\pmb{a},\pmb{\lambda}) = \sum_{ K  \subset M \times M } [\prod_{\{ (i,j) \in K^c \}} (1-r_{i,j}) \prod_{\{ (h,k) \in K \}} r_{h,k}]  W(\pmb{a}, (1-\rho)^{\phi(K,A)} \otimes \lambda),
\end{equation}
where $\otimes$ denotes multiplication coordinatewise.

\medskip

Equations (\ref{risk-welfare1}) and (\ref{risk-welfare2}) highlight that, in this extended setting, social welfare depend on the structure of the production network, the ``spatial" distribution of risks given by $r$, and the  disruption functions $\phi$. Socially efficient networks shall both minimize the risk of disruption and ensure resilience in case a disruption occurs. One can characterize more precisely these efficient networks once the distribution of risks and the disruption functions are specified.

In order to disentangle between risk and productivity related effects in the game $\mathcal{G}^n(a_0,b)$ where all firms have Hicks-neutral productivity functions, we further assume that utility weights are independent of the category/country, i.e. for all $\ell$ and for all $i,j \in M_\ell$, $a_{0,i}=a_{0,j}$. In this setting, we consider two polar cases for the disruption functions:
\begin{itemize}
\item Firstly, there is a fixed-cost to disruption independently of the number of links affected in each category. It is characterized by $\phi^{\min}$ such that for all $i \in M$,
\begin{equation*} 
\phi_i^{\min}(K,A)=\displaystyle\sum_{\ell \in L}  \min_{j \in M_{\ell} \mid (i,j) \in K \text{ and } a_{i,j} >0} a_{i,j}. 
\end{equation*}
\item Secondly, the impacts on the suppliers cumulate additively. It is characterized by $\phi^{\text{sum}}$ such that for all $i \in M$,
\begin{equation*} 
\phi^{\text{sum}}_i(K,A)= \displaystyle\sum_{j \mid (i,j) \in K \text{ and }  a_{i,j} >0} a_{i,j}. 
\end{equation*}
\end{itemize}
Furthermore, we consider two polar cases for the spatial distribution of risks:
\begin{itemize}
\item First, a case where risk is homogeneous, i.e. for all $i,j \in M$  $r_{i,j} = r$ for some $r \in [0,1].$
\item Second, a case where risk depends on the distance between firms, i.e. for $i \in M_{\ell}$ and $j\in M_{\ell'}$,  $r_{i,j}=r\frac{|\ell-\ell'|+1}{n}$ for some $r \in [0,1].$
\end{itemize}

In this setting, one can provide a complete comparative static for the $\mathcal{Q}$-clustered equilibria of the game $\mathcal{G}^n(\pmb{a}_0,\pmb{b})$. 
\begin{prop}\label{prop_risk} In the normal form game $\mathcal{G}^n(\pmb{a}_0,\pmb{b})$ with risk,
\begin{itemize}
\item[1)] If $\phi^i(K,A)=\phi^{\min}_i(K,A)$, then the maximum (resp. minimum) social welfare is achieved in the fully connected (resp. island) economy, irrespective of whether risk is homogeneous or increases with distance.
\item[3)] If $\phi^i(K,A)=\phi^{\text{sum}}_i(K,A)$ and risk is homogeneous, then the social welfare is the same for all $\mathcal{Q}$-clustered equilibria.
\item[3)] If $\phi^i(K,A)=\phi^{\text{sum}}_i(K,A)$ and risk is increasing with distance, then the maximum (resp. minimum) social welfare is achieved in the the island (resp. fully connected) economy.
\end{itemize}
\end{prop}

Despite the stylized nature of the examples considered, Proposition \ref{prop_risk} highlights a range of results relevant for applications. When supply disruptions are not cumulative (Case 1), diversification allows to mitigate risk and a fully connected network structure is always preferable, independently of the spatial distribution of risk. When supply disruptions cumulate additively and the spatial distribution of risks is homogeneous (Case 2), the benefits of diversification are exactly compensated by the increase costs of risks, so that every equilibrium network structure is equivalent. When supply disruptions cumulate additively and risk is proportional to spatial distance (Case 3), the benefits of diversification are more than offset by the increase in risk so that an autarkic organization is preferable.

\subsection{Welfare improving network policies}

The previous results of this section highlight that multiple production networks can emerge as strategic equilibria of decentralized network formations processes. Certain equilibria might be very inefficient from the social welfare point of view. In particular, firms may fail to account for the productivity gaps and/or the risks associated to certain supply relationships.

However, the clustered nature of these equilibria put forward in Proposition \ref{K-cluster} hint at relatively simple ways to restrict the set of possible equilibria in order to coordinate behavior on efficient equilibria. Indeed, in order to prevent the formation of a given cluster at equilibrium, it suffices to prevent the formation of a single link in that cluster. Conversely, to opt for certain clusters at equilibrium, it suffices to ensure \textit{\`a priori} that one of the links of the cluster will be formed. In practice, the creation or prevention of links can be enforced by a range of tools such as trade agreements, tariffs, and non-tariff barriers.

Formally, we define a trade policy by a pair of subgraphs $(\mathcal{P}, \mathcal{C}) \in M^2 \times M^2$ where $\mathcal{P}$ is the set of links that are ``prevented" by the policy
and  $\mathcal{C}$ the set of links that are ``catalyzed" by the policy. It is assumed that a prevented link can not be present in an equilibrium network while a subsidized link must be present in an equilibrium network. More precisely, we say that a $\mathcal{Q}$-clustered equilibrium network is compatible with a trade policy $(\mathcal{P}, \mathcal{S})$ if it contains none of the links in $\mathcal{P}$ and all the links in $\mathcal{S}.$ A trade policy can efficiently coordinate behavior on certain equilibria in the sense that, given the clustered nature of equilibria, one requires to prevent and/or catalyze only a few links to select specific equilibria. Namely, one has the following proposition: 

\begin{prop} \label{prop-policy} In any  $n$-fold replicate game $\mathcal{G}^n(\pmb{a}_0,\pmb{b}),$
\begin{itemize}
\item if a trade policy  $(\mathcal{P}, \mathcal{C})$ prevents at least one link from category $\ell$ to category $\ell',$ then category $\ell$ and category $\ell'$ are in separate clusters at every compatible $\mathcal{Q}$-clustered equilibrium network; 
\item if a trade policy  $(\mathcal{P}, \mathcal{C})$ catalyzes at least one link from category $\ell$ to category $\ell',$ then category $\ell$ and category $\ell'$ are the same cluster at every compatible $\mathcal{Q}$-clustered equilibrium network. 
\end{itemize}
\end{prop}

\section{Conclusion}\label{conclusion}

We have developed a strategic model of the formation of production networks.  Our models subsumes standard general equilibrium models as, once firms have chosen their supply relationships, the emerging outcome is a general equilibrium of the economy with the production structure determined  by the firms' choices. Accordingly, the objective of firms in the network formation game is to choose their supply relationships so as to maximize their profit at the resulting equilibrium. In other words, firms aim to insert themselves efficiently in the network of exchanges. We have also shown that the objective of the firms is equivalent to the maximization of their eigenvector centrality in the production network. 

As is common in network formation games based on centrality, there generally are multiple Nash equilibria in our setting. We have investigated the characteristics and the social efficiency of these equilibria in a stylized version of our model representing international trade networks. We show that the impact of network structure on social welfare is firstly determined by a trade-off between costs of increasing process complexity and positive spillovers on productivity induced by the diversification of the input mix. If the latter effect dominates, strongly interconnected, globalized, network structures are socially efficient. Conversely, if the cost from increasing process complexity dominates, island-based, localized, network structures are socially efficient. 

We further analyze a variant of our model that accounts for the risks of disruption of supply relationships. In this setting, we characterize how social welfare depends on the structure of the production network, the spatial distribution of risks, and the process of shock aggregation in supply chains.  Optimal network structure is hence dependent on the riskiness environment in which the economy operates. In particular, risky supply relationships might be profitable for the firm but generate substantial negative externalities at the macro-level. We finally show that simple trade policies characterized by sets of links that are either prevented or catalyzed can be a powerful equilibrium selection device. 
 
Our results highlight that the validity of standard welfare theorems is jeopardized when one subsumes general equilibrium models into a broader model of the formation of production networks including strategic considerations. From the theoretical point of view, our results are in line with the network theoretic literature that has shown that decentralized network formation processes could lead to inefficient macro-level outcomes in a range of socio-economic settings such as  polarization in social networks, epidemic spreading in human networks or systemic risk in financial networks. From the policy perspective, our results speak to the current debate about de-globalization by highlighting that the characteristics of  efficient production and trade networks might depend on  contextual factors such as environmental or geopolitical risks.

Our results are obtained in a highly stylized setting where all production functions are assumed to be Cobb-Douglas. In this specific case, firms are strategically ``neutral" about production costs because equilibrium revenues and profits are independent of productivity. This implies in particular that firms do not have incentives to minimize production costs as opposed to \cite{acemoglu2020endogenous}.Setting with alternative assumptions on elasticities of substitution could substantially modify our results. If intermediary inputs were perfect substitutes, firms would have incentives to target minimal production costs during the network formation step in order to increase their revenues in the resulting general equilibrium. Oppositely, if intermediary inputs were perfect complements, firms would have incentives to target maximal production costs during the network formation step in order to increase their revenues in the resulting general equilibrium. A detailed analysis of strategic network formation in these more complex environments and the consideration of intermediary  objectives between profit maximization and cost minimization  seem promising directions for future research.

\section*{Acknowledgments}

AM acknowledges financial support from the European Union's Horizon 2020 research and innovation programme under the grant agreement No 956107, ``Economic Policy in Complex Environments (EPOC)''.

\bibliographystyle{plainnat}

\bibliography{prodnet} 

\section*{Appendix A: Proofs} 

\begin{demo}[of Lemma  \ref{ergodic}]
Let us begin by noting that for all $i \in M$, $a_{i,0}=b_{i,0}>0$. Moreover, since $b_{i_0,\ell}>0$ for some $\ell \in L\setminus\{0\}$, there exists $j_0 \in M$ such that $a_{i_0,j_0} >0$. \color{black} The irreducibility of $\widetilde{A}$ is evident since, for all $i,j \in M,$ we have $a_{0,i} >0,$ $a_{j,0} >0$, and consequently, $a_{j,0} a_{0,i} >0$. The state $0$ is aperiodic due to the conditions $a_{0,i_0} a_{i_0,0} >0$ and $a_{0,i_0} a_{i_0,j_0} a_{j_0,0} >0$.
\end{demo}

\begin{demo}[of Proposition  \ref{prop-equi}]
Given a price vector $\pmb{p} \in \R^N_+$, the first-order condition for profit maximization by firm $i$ implies that, for all $j \in N$,
$$p_j x_{i,j}= a_{i,j} p_i \lambda_i \prod_{k\in N} x_{i,k}^{a_{i,k}}=a_{i,j} p_i y_i.$$
This implies that the output of firm $i$ is determined by
\begin{equation}
y_i:= f_{\pmb{a}_i}(x_{i,0},\cdots,x_{i,n})=\bigg[\lambda_i  p_i^{\sum_{j \in N} a_{i,j}} \prod_{j \in N} \bigg(\dfrac{a_{i,j}}{p_j}\bigg)^{a_{i,j}}\bigg]^{\dfrac{1}{1-\sum_{j \in N} a_{i,j}}}, \label{prod}
\end{equation}
and the profit of firm $i$ is expressed as
\begin{equation*} 
\pi_i:= p_iy_i(1-\sum_{j \in N} a_{i,j}). 
\end{equation*}

The first-order conditions for utility maximization yield that, for all $j \in M$,
$$p_j x_{0,j} = a_{0,j} p_{0} + a_{0,j} \sum_{i=1}^{n} p_i y_i (1-\sum_{k \in N} a_{i,k}).$$

The market clearing condition for good $j \in M$ is given by
$$y_j= \dfrac{ a_{0,j} p_{0} + a_{0,j} \sum_{i=1}^{n} p_i y_i (1-\sum_{k \in N} a_{i,k})}{p_j} +\sum_{k \in M} \dfrac{a_{k,j}p_ky_k}{p_j},$$
or equivalently,
\begin{equation}p_j y_j=  a_{0,j} p_{0} + a_{0,j} \sum_{i=1}^{n} p_i y_i (1-\sum_{k \in N} a_{i,k})+\sum_{k \in M} a_{k,j}p_ky_k. \label{eq-goodclear} \end{equation} Furthermore, the market clearing condition for the labor market yields
$$ 1 = \sum_{j \in M} \dfrac{ a_{j,0}p_j y_j}{p_0},$$
which can be expressed as
\begin{equation} p_0 = \sum_{j \in M} a_{j,0}p_j y_j. \label{eq-labclear} \end{equation}

Considering $j \in M$ and defining the revenue of firm $j$ as $v_j = p_j y_j$, Equations \eqref{eq-goodclear} and \eqref{eq-labclear} establish that, for every $j \in M$, the following relationships hold:
\begin{equation}
v_j = a_{0,j} v_0 + \sum_{k \in M} a_{k,j} v_k, \label{goodval}
\end{equation}
where $v_0$ represents household revenues defined by
\begin{equation} v_0= p_0+  \sum_{i \in M} \varepsilon_i v_i= \sum_{i \in M}( a_{i,0} + \varepsilon_i) v_i=  \sum_{i \in M}  (1-\sum_{k \in M} a_{i,k})v_i. \label{labval}
\end{equation}

Equations $\eqref{eq-labclear}$ and $\eqref{goodval}$ can be expressed in matrix form as
\begin{equation}
 \left(\begin{array}{c}p_0 \\\pmb{v} \end{array}\right) =\widetilde{A}^T \left(\begin{array}{c}p_0 \\\pmb{v} \end{array}\right), \label{eq-inv}
\end{equation}
Since $\widetilde{A}$ is row-stochastic, aperiodic, and irreducible, the Perron-Frobenius theorem straightforwardly implies the existence of $p_0 \in \R_+$ and $\pmb{v} \in \R^M_+$ such that Equation $\eqref{eq-inv}$ holds. Furthermore, $\pmb{v}$ is unique up to price normalization and is entirely determined by the choice of $p_0$ in Equation $\eqref{eq-labclear}$, which now becomes $\sum_{j \in M} a_{j,0} \overline{v}_j=1$. 

Finally, to demonstrate the existence and characterize equilibrium prices, Equation (\ref{prod}) is employed, yielding that, for all $i \in M$,
\begin{equation}
v_i = \left[\lambda_i p_i \prod_{j \in N} \left(\frac{a_{i,j}}{p_j}\right)^{a_{i,j}}\right]^{\frac{1}{1-\sum_{j \in N} a_{i,j}}}. \label{eq-prod2}
\end{equation}
By taking logarithms on both sides of Equation (\ref{eq-prod2}), we obtain
\begin{align*}
&\log(v_i)= \dfrac{1}{1-\sum_{j \in N} a_{i,j}} [\log(\lambda_i) + \log(p_i) + \sum_{j \in N} a_{i,j} \log(a_{i,j})  - \sum_{j \in N} a_{i,j} \log(p_j)] \\
\Longleftrightarrow &(1-\sum_{j \in N} a_{i,j})\log(v_i)=  [\log(\lambda_i) + \log(p_i) + \sum_{j \in N} a_{i,j} \log(a_{i,j})  - \sum_{j \in N} a_{i,j} \log(p_j)].
\end{align*}
By normalizing $p_0=1,$ the expression can be reformulated as
$$\sum_{j \in M} a_{i,j} \log(p_j) - \log(p_i)  =  \log(\lambda_i) + \sum_{j \in N} a_{i,j} \log(a_{i,j}) -(1-\sum_{j \in N} a_{i,j})\log(v_i),$$
which can be compactly represented in matrix form
$$(A-I) \log(\pmb{p})= u + D \log(\pmb{v}),$$
where $u_i= \log(\lambda_i)+  \sum_{j \in N} a_{i,j} \log(a_{i,j}$, $D=\diag( \sum_{j \in N} a_{i,j} -1)$, $\log(p)_i=\log(p_i)$ and $\log(v)_i=\log(v_i)$. Since, for all $i,$ $a_{i,0} > 0,$ it is evident that the spectral radius of $A$ is less than one, making $(A-I)$ invertible. Consequently, we can derive $$\log(\pmb{p})= (A-I)^{-1} \pmb{u}  +(A-I)^{-1} D \log(\pmb{v}).$$

\end{demo}

\begin{demo}[of Lemma \ref{prof2} ]

Using Equation (\ref{eq-goodclear}) and the fact that $p_0=1$, one obtains that for $ i \in M,$
$$v_j= a_{0,j}+ a_{0,j} \sum_{i\in N} v_i \varepsilon_i + \sum_{i \in M} a_{i,j} v_i,$$
which is equivalent to that
$$\pmb{v}= \pmb{a}_0+ \widetilde{A}_M \pmb{v}.$$

Now since $\widetilde{A}$ is row-stochastic and for all $i \in M$, $\tilde{a}_{i,0}>0$, it is clear that for all $i \in M$, the sum of all entries on column $i$ of matrix $\widetilde{A}_M$ is strictly less than 1. Thus, the largest eigenvalue of $\widetilde{A}_M$ is less than one, and consequently, $(I-\widetilde{A}_M)$ is invertible. One then has
\begin{equation*}
\pmb{v}=(I-\widetilde{A}_M)^{-1}\pmb{a}_0. 
\end{equation*}
Furthermore, we can write
\begin{equation*}
\pmb{v}= \sum_{n=0}^{+\infty} (\widetilde{A}_M)^{n}\pmb{a}_0,
\end{equation*}
which, in turn, yields
\begin{equation*}
\pmb{v}=  \sum_{j \in M} a_{0,j} P_{j,i}(\pmb{a}).
\end{equation*}
Finally, it is sufficient to substitute these expressions into $\pi_i(\pmb{a}) = \varepsilon_i v_i$ to conclude the proof.

\end{demo}

\begin{demo}[of Lemma  \ref{profchar}]

Following Lemma \ref{prof2}, one has $\pi_i(\pmb{a})= \varepsilon_i  \sum_{j \in M} a_{0,j} P_{j,i}(\pmb{a})$. It thus suffices to prove that for all $j \in M\setminus \{i\}$,

\begin{equation*} 
P_{j,i}(\pmb{a})= \dfrac{D_{j,i}(\pmb{a})}{1-D_{i,i}(\pmb{a})}
\end{equation*}
and
\begin{equation*} 
P_{i,i}(\pmb{a})= \dfrac{1}{1-D_{i,i}(\pmb{a})}.
\end{equation*}

\noindent For any path $p \in \mathcal{P}_{j,i}$ with $j \not=i$, it is clear that either the path is a direct path or it can be (uniquely) decomposed into a directed path $d \in \mathcal{D}_{j,i}$ and a cycle $c$ around $i$. We use $\mathcal{C}_{i}$ to denote the set of all cycles around $i$. 

Then, it is straightforward to get
\begin{equation*}
P_{j,i}(\pmb{a})= D_{j,i}(\pmb{a}) (1+ C_i(\pmb{a})), 
\end{equation*}
where $C_i(\pmb{a}):= \sum_{c \in \mathcal{C}_{i}} w_a(c)$ is the sum of weights of cycles around $i$.

Now, let us define the multiplicity of a cycle around $i$. We say that a cycle around $i$ has a multiplicity of $k\geq 1$ if node $i$ appears $k+1$ times in the cycle. We denote the set of all cycles around $i$ of multiplicity $k$ as $\mathcal{C}^k_i$ and remark that $\mathcal{C}^1_i=\mathcal{D}_{i,i}$. We note that $\mathcal{C}_{i} = \cup_{k=1}^{\infty} \mathcal{C}^k_i,$ i.e. every cycle around $i$ of multiplicity $k$ can be decomposed into $k$ cycles around $i$ of multiplicity 1, and the decomposition is unique. This implies that
\begin{equation*}  
C_i(\pmb{a}) = \sum_{k=1}^{+\infty} (D_{i,i}(\pmb{a}))^k.
\end{equation*}
Thus,
\begin{equation*}
P_{j,i}(\pmb{a})  =  D_{j,i} (\pmb{a})(1+ C_i(\pmb{a})) =  D_{j,i}(\pmb{a})  \sum_{k=0}^{+\infty} (D_{i,i}(\pmb{a}))^k = \dfrac{D_{j,i}(\pmb{a})}{1-D_{i,i}(\pmb{a})}, 
\end{equation*}
where one has used Remark \ref{markovrem} which ensures that $D_{i,i} \leq P_{i,i}<1$. A similar reasoning shows that
\begin{equation*}
P_{i,i} = \dfrac{1}{1-D_{i,i}}. 
\end{equation*}

Finally, the linearity of $1-\tilde{a}_{i,i}-\sum_{k \in M\setminus\{i\}} \tilde{a}_{i,k} D_{k,i}(\pmb{a} )$ in $\tilde{a}_{\cdot,i}$, and thus in $a_{\cdot,i}$, implies that the profit of firm $i$ is continuous and quasi-concave in $\pmb{a}_i$. 
\end{demo}

\begin{demo} [of Proposition \ref{prop-nash}]
One considers the equivalent game where the payoff of agent $i$ is given by the invariant probability of $i \in M,$ $\pi_i,$ for the Markov chain defined by $\widetilde{A}.$ Now, following the Markov chain tree theorem (\cite[see e.g.][]{kemeny1976finite}), $\pi_i$ can be characterized as follows:
\begin{itemize}
\item For each $i \in M,$ denote $T_i$ as the set of $i$-rooted spanning trees. These trees are defined as acyclic graphs with a node set of $M$, where node $i$ has no outgoing edges, and every other node $j \in M\setminus \{i\}$ possesses an out-degree of $1$.
\item The weight of a tree $t \in T_i$ is defined as $w(t):= \prod_{(j,k) \in t} \widetilde{A}_{j,k}$, where the notation $(j,k)$ denotes the edges of $t$. Notably, $j \neq i$ always holds, ensuring that $w(t)$ remains independent of $a_i$ for $t \in T_i$.
\item The weight of $T_i$ can then be defined as $w(T_i):= \sum_{t \in T_i} w(t)$, and $w_i$ is given by
\begin{equation*} 
\pi_i:= \dfrac{w(T_i)}{\sum_{j \in M} w(T_j)}.
\end{equation*}
\end{itemize} 
Since $w(T_i)$ is independent of $a_i$, each agent $i$ seeks to maximize $\Phi(\pmb{a}):=\dfrac{1}{\sum_{j \in M} w(T_j)}$. Notice that this function is an ordinal potential function for the game, which ends the proof.
\end{demo}

\begin{demo} [of Proposition \ref{ownsupply}]
It follows from Lemma \ref{profchar} that, at a Nash Equilibrium, profit maximization amounts to maximize 
$$\tilde{a}_{i,i} +\sum_{k \in M \setminus \{i\}} \tilde{a}_{i,k} D_{k,i}(\pmb{a} )= a_{i,i} + \varepsilon_i a_{0,i} +\sum_{k \in M\setminus\{i\}} (a_{i,k} +\varepsilon_i a_{0,k}) D_{k,i}(\pmb{a}).$$
Furthermore, by Remark \ref{markovrem}, one necessarily has  $D_{k,i}(\pmb{a})\leq P_{k,i}(\pmb{a})<1$. In this setting, it is straightforward that if $\overline{\pmb{a}}_i$ maximizes profit given $\overline{\pmb{a}}_{-i}$, then $\a_{i,i}$ must be as large as possible, i.e. one must have $ \a_{i,i} = b_{i,\ell}$.
\end{demo}

\begin{demo}[of Proposition \ref{prop-firstorderwelfare}]
It suffices to show that $$\dfrac{\partial W(A,\pmb{\lambda})}{ \partial a_{i,j}} (\pmb{a})=(\pmb{a}_0)^T [  (I-A)^{-1} U_{i,j} (I-A)^{-1}]\pmb{u}+(\pmb{a}_0)^T(I-A)^{-1}\pmb{u}_{i,j},$$ where $\pmb{u}_{i,j}$ is a column vector 
 $$\begin{pmatrix}
 0\\
 \vdots\\
 0\\
 \log(a_{i,j})+1\\
 0\\
 \vdots\\
 0
 \end{pmatrix}.$$
The proposition follows by expressing the coefficients of $(I-A)^{-1} =\sum_{n=0}^{+\infty} A^n$ in terms of network paths and applying Karush-Kuhn-Tucker conditions.

\end{demo}

Remark that, in the $\mathcal{Q}$-cluster economy, for every $j, j' \in M_{\ell'}$ and for every input $\ell$, $b_{j,\ell}=b_{j',\ell}$. Therefore, from now on, in the $\mathcal{Q}$-cluster economy, we denote $b_{\ell',\ell}$ the input weight  $b_{j,\ell}$ for some firm $j\in M_{\ell'}$.  To simplify of the notation, let us denote $\mathcal{Q}_k= \bigcup_{\ell \in L} Q_{\ell,k}$ the set of all firms in the cluster $Q_k$.

\begin{demo}[of Proposition \ref{K-cluster}]

Consider $\mathcal{Q}=\{Q_1,\ldots,Q_K\}$ and the associated vector of input weights for firm $i$, $a_i^\mathcal{Q}$ defined in System (\ref{clusters}). We aim to prove that for every firm $i$, an optimal vector of input weights is $\pmb{a}_i^\mathcal{Q}$ provided that the vector of input weights of all other firm $j$ is $\pmb{a}_j^\mathcal{Q}$. From now on, in this proof, we substract the superscript $\mathcal{Q}$ in the weight $\pmb{a}_i$ for simplification.

Let us consider firm $i$ producing goods of type $\ell$ in a cluster $Q_k$, i.e. $i \in Q_{\ell,k}$. Applying Proposition \ref{ownsupply}, the optimal weights for the input of type $\ell$ is $a_{i,i}=b_{\ell,\ell}$, $a_{i,j}=0$ for all $j \in M_\ell$ and $j\neq i$.

For input of type $\ell' \neq \ell$, let us look at the profit function of firm $i$ at the vector of input weights $\pmb{a}=(\pmb{a}_i,\pmb{a}_{-i})$, which is given by Lemma \ref{profchar}
\begin{align*}
\pi_i(\pmb{a}_i,\pmb{a}_{-i})=\varepsilon_i \dfrac{a_{0,i} +\sum_{j \in M \setminus\{i\}} a_{0,j} D_{j,i}(\pmb{a})}{1-\tilde{a}_{i,i}-\sum_{k \in M\setminus\{i\}} \tilde{a}_{i,j} D_{j,i}(\pmb{a} )}.
\end{align*} 
Recall that $D_{j,i}(\pmb{a})$ is the sum of weights of all direct paths from $j$ to $i$ in the adjusted matrix $\widetilde{A}_M$ defined in Section \ref{Net_based_equi}. There are two possible cases:

\medskip

Case 1: $j \in Q_{\ell',k}$. We have $D_{j,i}(\pmb{a}) \geq \tilde{a}_{i,j} \geq \dfrac{b_{\ell',\ell}}{|Q_k|}$.

\medskip

Case 2: $j \notin Q_{\ell',k}$. We remark that, for any $i,j \in M$, the sum of weights of all paths from $j$ to $i$ is at most  at most $C=\displaystyle\max_{i \in M} \frac{1}{1-b_{i,0}}>0$. Now, note that any path $p$ from $j$ to $i$ can be presented as follows:
$$\begin{cases}
p = (j,h_1,\dots ,h_q,h_{q+1}\dots,i),\\
h_{q'}\notin \mathcal{Q}_k \text{ for all } q'\in \{1,\ldots,q-1\}, h_q \in \mathcal{Q}_k
\end{cases}$$
Intuitively, the first part of $p$, i.e. $(j,h_1,\dots ,h_q)$, contains firms which are not in the clusters $Q_k$ except the terminal firm $h_q$. If we denote $D_{j,h_q}(\pmb{a},Q_k)$ the sum of weights of such direct paths from $j$ to $i$, then we have
$$D_{j,h_q}(\pmb{a},Q_{k})=\sum_{j' \notin \mathcal{Q}_k} D_{j,j'}(\pmb{a})\tilde{a}_{j',h_q}.$$
Note that, since $j' \notin \mathcal{Q}_k$ and $h_q \in \mathcal{Q}_k$ one gets $\tilde{a}_{j',h_q}= \varepsilon_{h_q} a_{0,j'} \leq D=\displaystyle\max_{i\in M}a_{0,i} \times \displaystyle\max_{i\in M} \varepsilon_i$ . Moreover, there are at most $(n-|Q_k|) \times L$ such $j'$, which follows
$$D_{j,h_q}(\pmb{a},Q_k)\leq C \times (n-|Q_k|) \times L \times D.$$
Therefore, one obtains 
\begin{align*}
D_{j,i}(\pmb{a})=\sum_{h_q \in \mathcal{Q}_k}D_{j,h_q}(\pmb{a},Q_{k})\times D_{h_q,i}(\pmb{a})&\leq \big(C \times (n-|Q_k|) \times L \times D\big) \times \big(n\times |Q_k| \times C\big)\\
&<C^2 \times \frac{n^3}{2} \times L \times D.
\end{align*}
\medskip

Note that $C$, $n$, and $L$ are fixed, and $D$ converges to  $0$ when $\max_{i\in M} \varepsilon_i \to 0$. Therefore, $D_{j,i}(\pmb{a}) \to 0$ when $\max_{i\in M} \varepsilon_i \to 0$. So, if $ \varepsilon_i$ is small enough for all $i$, then $D_{j_1,i}(\pmb{a}) > D_{j_2,i}(\pmb{a})$ for any $j_1\in Q_{\ell',k}$ and $j_2 \notin Q_{\ell',k}$. Therefore, the optimal weights of firm $i$ to firm $j \notin Q_{\ell',k}$ is $0$. \color{black}

If $j_1, j_2 \in Q_{\ell',k}$, then $D_{j_1,i}(\pmb{a})=D_{j_2,i}(\pmb{a})$ due to the fact that $j_1$ and $j_2$ have the same vector of input weights since they are replication of each other. Then $a_{i,j}=\frac{b_{\ell,\ell'}}{|Q_k|}$ for all $j \in Q_{\ell',k}$ are clearly optimal weights for input type $\ell'\neq \ell$, which ends of the proof.
\end{demo}

The proofs of  Propositions \ref{prop_hick_neutral} and \ref{prop_diversification} rely on the following Lemma \ref{inverse_matrix}, which characterizes the Leontief inverse matrix $(I-A)^{-1}$.

\begin{lemma}\label{inverse_matrix} Let $A^{Q_k}$ be the production network matrix associated to the network structure $a^\mathcal{Q}$ restricted to the cluster $Q_k$ and define $(I-A^{Q_k})^{-1}=C^{Q_k}=[c_{i,j}]_{(|Q_k| L)\times (|Q_k| L)}$, then for $i \in Q_{\ell,k}$, one has
 $$
\begin{cases}
c_{i,i}=\frac{C_{\ell,\ell}}{|Q_k|}+\frac{|Q_k|-1}{|Q_k|}\frac{1}{1-b_{\ell,\ell}}, \\
c_{i,j}=\frac{C_{\ell,\ell}}{|Q_k|}-\frac{1}{|Q_k|}\frac{1}{1-b_{\ell,\ell}} \text{ for }  j \in Q_{\ell,k}\setminus \{i\}, \\
c_{i,j}=\frac{C_{\ell,\ell'}}{|Q_k|} \text{ for } j  \in Q_{\ell',k} \text{ with } \ell' \neq  \ell,
\end{cases}$$
where $C_{\ell,\ell'}$ does not depend on the choice of $\mathcal{Q}$-cluster.
\end{lemma}

\begin{demo}[of Lemma \ref{inverse_matrix}]

Let $|Q_k|=m_k$, and let $I-A^{Q_k}=[d_{i,j}]_{m_kL \times m_kL}$. Considering $i \in Q_{\ell,k}$, we have
\begin{align}
\displaystyle\sum_{j\in \mathcal{Q}_k}d_{i,j}c_{j,h}=\begin{cases}
0 \text{ if } i\neq h,\\
1 \text{ if } i =h.
\end{cases}
\label{eq_cd}
\end{align}
Now, for all $\tilde \ell$, we denote $C_{i,\tilde\ell}=\displaystyle\sum_{j \in Q_{\tilde\ell,k}} c_{i,j}$. Summing up Equation (\ref{eq_cd}) over $h \in Q_{\ell,k}$ and over $h \in Q_{\ell',k}$ for every $ \ell' \neq  \ell$, one gets
\begin{align*}
\displaystyle\sum_{h \in Q_{\ell,k}}\left(\displaystyle\sum_{j\in \mathcal{Q}_k}d_{i,j}c_{j,h}\right)=1,\\
\displaystyle\sum_{h \in Q_{\ell',k}}\left(\displaystyle\sum_{j\in  \mathcal{Q}_k}d_{i,j}c_{j,h}\right)=0
\end{align*}
By interchanging the indices, this is equivalent to that
\begin{align}
\displaystyle\sum_{j\in  \mathcal{Q}_k}d_{i,j}C_{j,\ell'}=\begin{cases}
0 \text{ if } \ell' \neq \ell,\\
1 \text{ if } \ell'=\ell
\end{cases}
\label{eq_cd_2}
\end{align}
Since $i \in M_{\ell}$, one has
\begin{align*}
d_{i,i}&=1-b_{\ell,\ell}\\
d_{i,j}&=0 \text{ for } j \in M_\ell \setminus \{i\}\\
d_{i,j}&=-\frac{b_{\ell,\ell'}}{m_k} \text{ for } j \in Q_{\ell',k} \text { with } \ell \neq \ell'
\end{align*}
 Equation (\ref{eq_cd_2}) can be written as follows
\begin{align*}
(1-b_{\ell,\ell})C_{i,\ell'}+\displaystyle\sum_{\tilde{\ell}\neq \ell}\left(\displaystyle\sum_{j \in Q_{\tilde{\ell},k}}\left(-\frac{b_{\ell,\tilde{\ell}}}{m_k}\right)C_{j,\ell'}\right)&=0, \text{ for all } \ell' \neq \ell,\\
(1-b_{\ell,\ell})C_{i,\ell}+\displaystyle\sum_{\tilde{\ell}\neq \ell}\left(\displaystyle\sum_{j \in Q_{\tilde{\ell},k}}\left(-\frac{b_{\ell,\tilde{\ell}}}{m_k}\right)C_{j,\ell}\right)&=1
\end{align*}
This implies that, for all $i,j \in Q_{\ell,k}$ and for all $\ell'$, $C_{i,\ell'}=C_{j,\ell'}$. Thus, for $i \in Q_{\ell,k}$, we can denote $C_{i,\ell'}=C_{\ell,\ell'}$. Equation (\ref{eq_cd_2}) now becomes
\begin{align*}
(1-b_{\ell,\ell})C_{\ell,\ell'}+\displaystyle\sum_{\tilde{\ell}\neq \ell}\left(-b_{\ell,\tilde{\ell}}C_{\tilde{\ell},\ell'}\right)&=0, \text{ for all } \ell' \neq \ell,\\
(1-b_{\ell,\ell})C_{\ell,\ell}+\displaystyle\sum_{\tilde{\ell}\neq \ell}\left(-b_{\ell,\tilde{\ell}}C_{\tilde{\ell},\ell}\right)&=1.
\end{align*}
Denote $\pmb{C}_\ell=(C_{\ell',\ell})_{\ell' \in L} \in \mathbb{R}^L$ and $D=I-B$ where $B=[b_{\ell',\ell}]_{L \times L}$. Equations above implies that
$$D(\pmb{C}_\ell)^T=(\pmb{e}_\ell)^T$$
where $\pmb{e}_\ell = (0,\ldots,0,1,0,\ldots,0) \in \mathbb{R}^L$ is the $\ell$-th canonical basis. So $\pmb{C}_\ell=\pmb{e}_\ell (D^T)^{-1}$, which does not depend on the choice of $\mathcal{Q}$-cluster economy. Note that due to symmetry of cluster $Q_k$, it's easy to see that $c_{i,j}=c_{i,j'}$ for all $i,j,j'$ such that $i \in Q_{\ell,k}$, $j, j' \in Q_{\ell',k}$ with $\ell \neq \ell'$. Therefore, one obtains
$$c_{i,j}=\dfrac{C_{ \ell,\ell'}}{m_k} \text{ for } j  \in Q_{\ell',k} \text{ with } \ell' \neq  \ell$$
Besides, we have
\begin{align*}
\displaystyle\sum_{j\in \mathcal{Q}_k}c_{i,j}d_{j,h}=\begin{cases}
0 \text{ if } i\neq h,\\
1 \text{ if } i =h.
\end{cases}
\end{align*}
Taking $h \in Q_{\ell,k}\setminus \{i\}$ and $h=i$, we have
\begin{align*}
\displaystyle\sum_{\tilde{\ell}\neq \ell}\left(\displaystyle\sum_{j \in Q_{\tilde{\ell},k}} \frac{C_{\ell,\tilde{\ell}}}{m_k}\left(-\frac{b_{\tilde{\ell},\ell}}{m_k}\right)\right)+c_{i,h}(1-b_{\ell,\ell})&=0,\\
\displaystyle\sum_{\tilde{\ell}\neq \ell}\left(\displaystyle\sum_{j \in Q_{\tilde{\ell},k}} \frac{C_{\ell,\tilde{\ell}}}{m_k}\left(-\frac{b_{\tilde{\ell},\ell}}{m_k}\right)\right)+c_{i,i}(1-b_{\ell,\ell})&=1\\
\end{align*}
Therefore, $c_{i,i}-c_{i,h}=1/(1-b_{\ell,\ell})$ for all  $h \in Q_{\ell,k}\setminus \{i\}$. So $c_{i,h}=c_{i,h'}$ for all  $h,h' \in Q_{\ell,k}\setminus \{i\}$. Combining with $\sum_{h \in Q_{\ell,k}}c_{i,h}=C_{\ell,\ell}$, one gets
 $$
\begin{cases}
c_{i,i}=\frac{C_{\ell,\ell}}{m_k}+\frac{m_k-1}{m_k}\frac{1}{1-b_{ \ell, \ell}} \\
c_{i,j}=\frac{C_{\ell,\ell}}{m_k}-\frac{1}{m_k}\frac{1}{1-b_{ \ell, \ell}} \text{ for }  j \in Q_{\ell,k}\setminus \{i\} \\
\end{cases}$$
\end{demo}

\begin{demo}[of Proposition \ref{prop_hick_neutral}]
Let $\mathcal{Q}=\{Q_1,\ldots,Q_K\}$ be a partition of $\{1,\ldots,n\}$. Matrix $(I-A)^{-1}$ can be described as follows
\begin{equation*}
\begin{pmatrix}
C^{Q_1}&&&&\\
&\ddots&&&\\
&&C^{Q_i}&\\
&&&\ddots&\\
&&&&C^{Q_K}
\end{pmatrix}
\end{equation*}
Thus, we have for $i \in Q_{\ell,k}$,
\begin{align*}
\left((\pmb{a}_0)^T(I-A)^{-1}\right)_i&=\displaystyle\sum_{h \in M}a_{0,h}c_{h,i}\\
&=\displaystyle\sum_{\ell' \in L}\left(\displaystyle\sum_{h \in Q_{\ell',k}}a_{0,h}c_{h,i}\right)\\
&=\displaystyle\sum_{\ell' \in L}a_{0,\ell'}\left(\displaystyle\sum_{h \in Q_{\ell',k}}c_{h,i}\right) \text{ since } a_{0,h}=a_{0,h'}=a_{0,\ell'} \forall h, h' \in Q_{\ell',k}
\end{align*}
Now, by Lemma \ref{inverse_matrix}, one gets that $\sum_{h \in Q_{\ell',k}}c_{h,i}$ equals also $C_{\ell',\ell}$. Then we have
\[\left((\pmb{a}_0)^T(I-A)^{-1}\right)_i = \displaystyle\sum_{\ell' \in L}a_{0,\ell'}C_{\ell',\ell} \]
which is independent of the choice of $\mathcal{Q}$. 

If for all $i$, $\lambda_i=\bar \lambda$, then  $u_i=u_j=a_{i,0}\log a_{i,0}$ which is also independent of the structure of network. So the welfare $W(A,\pmb{\lambda})=(\pmb{a}_0)^T(I-A)^{-1} \pmb{u}$ are the same for all $\mathcal{Q}$-cluster economy.

\end{demo}

\begin{demo}[of Proposition  \ref{prop_diversification}]
We first show that, among all the $\mathcal{Q}$-clustered network equilibria, $\overline{\lambda}(a_i)$ achieves the highest (resp., lowest) value in the fully connected (resp., islands) economy. Indeed, assume that $\mathcal{Q}=\{Q_1,\ldots,Q_K\}$ and $i \in Q_{\ell,k}$. One gets
$$\log(\overline{\lambda}(\pmb{a}_i))=-\sum_{j \in Q_k}a_{i,j}\log(a_{i,j}).$$
Replacing $a_i$ defined in System (\ref{clusters}), we have
\begin{align*}
\log(\overline{\lambda}(\pmb{a}_i))=-b_{\ell,\ell}\log(b_{\ell,\ell})-\displaystyle\sum_{\ell' \neq \ell}b_{\ell,\ell'}\log\left(\frac{b_{\ell,\ell'}}{|Q_k|}\right).
\end{align*}
It is easy to see that, if $0<c<1$, function $x\log(c/x)$ is decreasing in the domain of $x\geq 1$. Thus, the value of $b_{\ell,\ell'}\log(b_{\ell,\ell'}/|Q_k|)$ is highest (resp., lowest) when $|Q_k|=1$ (resp., $|Q_k|=n$). Therefore, $\overline{\lambda}(\pmb{a}_i)$ achieves the highest (resp., lowest) value when $|Q_k|=n$ (resp., $|Q_k|=1$).

If we denote $\pmb{a}_i^F$ (resp., $\pmb{a}_i^I$) the vector of input weights of firm $i$ in the fully connected (resp., islands) economy, then $ \overline{\lambda}(\pmb{a}^F_i)  > \overline{\lambda}(\pmb{a}^I_i)$. Thus, if all firms have increasing returns to diversification, then one has
$$ \dfrac{\lambda_i(\pmb{a}^F_i)}{\overline{\lambda}(\pmb{a}^F_i)}> \dfrac{\lambda_i(\pmb{a}^I_i)} {\overline{\lambda}(\pmb{a}^I_i)}.$$
Now, note that for all $i$, $u_i= \log(\lambda_i(\pmb{a}_i))-\log(\overline{\lambda}(\pmb{a}_i))+b_{\ell,0} \log(b_{\ell,0})$. Therefore,
\begin{align*}
u^F_i&= \log(\lambda_i(\pmb{a}^F_i))-\log(\overline{\lambda}(\pmb{a}^F_i))+b_{\ell,0} \log(b_{\ell,0})>u^I_i \\
&= \log(\lambda_i(\pmb{a}^I_i))-\log(\overline{\lambda}(\pmb{a}^I_i))+b_{\ell,0} \log(b_{\ell,0}).
\end{align*}

Applying the similar computation as in Proof of Proposition \ref{prop_hick_neutral}, the welfare of the fully connected economy $W(A^F,\lambda)=(\pmb{a}_0)^T(I-A_F)^{-1} \pmb{u}^F$ is greater than the welfare of the islands economy $W(A^I,\lambda)=(\pmb{a}_0)^T(I-A_I)^{-1} \pmb{u}^I$, where $A^F$ (resp., $A^I$) is the production network associated to the fully connected (resp., islands) economy. The result for decreasing returns to diversification proceeds in the same way.
\end{demo}

\begin{demo}[of Proposition  \ref{poa-example}] Defining $\sum_{\ell' \in L}a_{0,\ell'}C_{\ell',\ell}= Q_\ell$, same computation as in the proof of Proposition \ref{prop_hick_neutral} gives that for all $i \in M_\ell$,
\begin{align*}
\left((\pmb{a}_0)^T(I-A)^{-1}\right)_i=\dfrac{1}{n}Q_\ell.
\end{align*}
The welfare $W(A,\pmb{\lambda})=(\pmb{a}_0)^T(I-A)^{-1} \pmb{u}$ can be written as follows:
\begin{align*}
W(A,\pmb{\lambda})&=\displaystyle\sum_i\left((a_0)^T(I-A)^{-1}\right)_i u_i\\
&=\displaystyle\sum_{\ell \in L}\dfrac{1}{n}Q_\ell\left(\displaystyle\sum_{i \in M_\ell} u_i \right).
\end{align*}
In the fully connected economy, we have for all $i \in M_\ell$,
$$u_i^F= \log(\lambda)+b_{\ell,0} \log(b_{\ell,0})+b_{\ell,\ell}\log(b_{\ell,\ell})+\displaystyle\sum_{\ell' \neq \ell}b_{\ell,\ell'}\log\left(\frac{b_{\ell,\ell'}}{n}\right).$$
In the islands economy, we have for all $i \in M_\ell$,
$$u_i^I= \log(\lambda)+b_{\ell,0} \log(b_{\ell,0})+b_{\ell,\ell}\log(b_{\ell,\ell})+\displaystyle\sum_{\ell' \neq \ell}b_{\ell,\ell'}\log\left(b_{\ell,\ell'}\right).$$
Thus, the difference between the welfare of the fully connected economy and the islands one is
\begin{align*}
W(A^I,\lambda)-W(A^F,\lambda)&=\displaystyle\sum_{\ell \in L}\dfrac{1}{n}Q_\ell\left(\displaystyle\sum_{i \in M_\ell} (u^I_i-u^F_i) \right)\\
&=\displaystyle\sum_{\ell \in L}Q_\ell\left(\displaystyle\sum_{\ell' \neq \ell}b_{\ell,\ell'}\log(n) \right)\\
&=\log(n) \displaystyle\sum_{\ell \in L}Q_\ell\left(1-b_{\ell,\ell}-b_{\ell,0} \right).
\end{align*}
It is clear that $Q_\ell\left(1-b_{\ell,\ell}-b_{\ell,0}\right) >0$ for all $\ell$. Thus denoting $K=\sum_{\ell \in L}Q_\ell\left(1-b_{\ell,\ell}-b_{\ell,0} \right)>0$ ends the proof.

\end{demo}

\begin{demo}[of Proposition \ref{prop_risk}]

The following lemmas are useful for the proof:
\begin{lemma} \label{lem_proba} Let $p_1,\ldots,p_n$ are real numbers. Then 
$$\sum_{k=1}^{n}k\left(\sum_{1\le i_1<\ldots<i_k\leq n}p_{i_1}\cdots p_{i_k}\prod_{j \neq i_1,\ldots,i_k}(1-p_j)\right)=\sum_{j=1}^n p_j.$$
\end{lemma}

\begin{demo}[of Lemma \ref{lem_proba}]
Denoting $N=\{1,\ldots,n\}$, we have for every $1\leq i \leq n$,
\begin{align*}
p_i&=p_i\sum_{k=0}^{n-1}\left(\sum_{\substack{1\le i_1<\ldots<i_k\leq n \\ i_1,\ldots,i_k \in N \setminus \{i\}}}p_{i_1}\cdots p_{i_k}\prod_{\substack{j \neq  i_1,\ldots,i_k \\j, i_1,\ldots,i_k \in N  \setminus \{i\}}}(1-p_j)\right)\\
&=\sum_{k=0}^{n-1}\left(\sum_{\substack{1\le i_1<\ldots<i_k\leq n \\ i_1,\ldots,i_k \in N  \setminus\{i\}}}p_ip_{i_1}\cdots p_{i_k}\prod_{\substack{j \neq  i_1,\ldots,i_k \\j, i_1,\ldots,i_k \in N  \setminus\{i\}}}(1-p_j)\right).
\end{align*}
By summing up over $i$, one can obtain the desired result by noting that, for fixing $1\le i_1<\ldots<i_k\leq n$, the term $p_{i_1}\cdots p_{i_k}\prod_{j \neq i_1,\ldots,i_k}(1-p_j)$ appears exactly $k$ times.
\end{demo}

\begin{lemma} \label{lem_sum_distance} Let $a_1,\ldots,a_n,b_1\ldots,b_m$ be real numbers. Then
\begin{equation}
\dfrac{\displaystyle\sum_{1\leq i\neq j \leq n}|a_i-a_j|}{n}+\frac{\displaystyle\sum_{1\leq i\neq j \leq m}|b_i-b_j|}{m}\leq \dfrac{\displaystyle\sum_{\substack{1\leq i \leq n \\1\leq j \leq m}}|a_i-b_j|}{m+n}.
\label{sum_distance}
\end{equation}
\end{lemma}

\begin{demo}[of Lemma \ref{lem_sum_distance}]
To prove Inequality (\ref{sum_distance}), we will prove the following equivalent inequality:
\begin{equation*}
\frac{m}{n}\left(\displaystyle\sum_{1\leq i\neq j \leq n}|a_i-a_j|\right)+\frac{n}{m}\left(\displaystyle\sum_{1\leq i\neq j \leq m}|b_i-b_j|\right)\leq\displaystyle\sum_{\substack{1\leq i \leq n \\1\leq j \leq m}}|a_i-b_j|.
\end{equation*}
Without loss of generality, we assume that all points belong to the segment $[0,1]$ and some two of the points are endpoints of the segment. Denote $A=\{a_1,\ldots,a_n\}$, $B=\{b_1,\ldots,b_m\}$, and $x=\{a_1,\ldots,a_n,b_1\ldots,b_m\}$.

Define function $f: [0,1]^{m+n} \to \mathbb{R}$ as follows:
$$f(x)=\frac{m}{n}\left(\displaystyle\sum_{1\leq i\neq j \leq n}|a_i-a_j|\right)+\frac{n}{m}\left(\displaystyle\sum_{1\leq i\neq j \leq m}|b_i-b_j|\right) -\left( \displaystyle\sum_{\substack{1\leq i \leq n \\1\leq j \leq m}}|a_i-b_j|\right).$$
It is clear that $f$ is a continuous function on a compact set, so it attains the minimum at some points $x$. Consider such a point with the smallest number of different values of coordinates. Then these different values belongs to $\{0,1\}$.

Indeed, assume that there is a group of coordinates whose value is strictly between 0 and 1. We remark that if we simultaneously increase (or decrease) these coordinates in its small neighborhood containing no other values of coordinates of $x$, the value of $f$ changes monotonically. Therefore, if we move this group of coordinates in the right direction which does not increase the value of $f$, we reach to other point $x'$, at which we attain minimum of $f$ and the number of different values of coordinates of $x'$ is smaller. This is a contradiction. 

So there is a minimum point $x^*$ of $f$ where there are $p$ numbers from $A$ and $q$ numbers from $B$ equal to $0$, and $n-p$ numbers from $A$ and $m-q$ numbers from $B$ equal to $1$. At $x^*$, we have
\begin{align*}
f(x^*)&=\dfrac{m}{n}p(n-p)+\dfrac{n}{m}q(m-q)-\left( p(m-q)+q(n-p) \right)\\
&=-\dfrac{(mp-nq)^2}{mn}\leq 0,
\end{align*}
which concludes the proof.
\end{demo}

Now, let us proceed with the proof of Proposition \ref{prop_risk}. We first observe that we can write the disruption functions $\phi^{min}_i(K,A)$ and $\phi^{sum}_i(K,A)$ as follows:
\begin{align*}
\phi^{\min}_i(K,A)&=\displaystyle\sum_{\ell \in L} \phi^{\min}_{i,\ell}(K,A),\\
\phi^{sum}_i(K,A)&=\displaystyle\sum_{\ell \in L} \phi^{sum}_{i,\ell}(K,A),
\end{align*}
where 
\begin{align*}
\phi_{i,\ell}^{\min}(K,A)&= \displaystyle\min_{j \in M_{\ell} \mid (i,j) \in K \text{ and }  a_{i,j} >0} a_{i,j}, \text{ and }\\
\phi^{\text{sum}}_{i,\ell}(K,A)&= \displaystyle\sum_{j \in M_{\ell} \mid (i,j) \in K \text{ and }  a_{i,j} >0} a_{i,j}.
\end{align*}
At the moment, we drop the superscript and write 
$$\phi_i(K,A)=\displaystyle\sum_{\ell \in L} \phi_{i,\ell}(K,A).$$
Since the productivity factor $\lambda_i$ is Hick-neutral, the expected (log) social welfare becomes
$$\text{ Const } + (a_0)^T(I-A)^{-1} u^r,$$
where
\begin{align*}
u^r_i&= \log(1-\rho) \displaystyle\sum_{ K  \subset M \times M } \left[\prod_{\{ (a,b) \in K^c \}}  (1-r_{a,b}) \prod_{\{ (c,d) \in K \}} r_{c,d}\phi_i(K,A)\right]\\
&=\log(1-\rho) \displaystyle\sum_{\ell \in L} \left(\displaystyle\sum_{ K  \subset M \times M } \left[\prod_{\{ (a,b) \in K^c \}} (1-r_{a,b}) \prod_{\{ (c,d) \in K \}} r_{c,d}\phi_{i,\ell} (K,A)\right]\right).
\end{align*}

Assume that $\mathcal{Q}=\{Q_1,\ldots,Q_K\}$. As computed below, the term $\phi_{i,\ell}(K,A)$ depends only on the cardinal of the set $K \cap \{(i,j)\mid j \in Q_{\ell,k}\}$, not on the set $K$ itself, because of the special structure of the $\mathcal{Q}$ cluster economy. Moreover, the links $(i,j)$ for $j \in M_\ell \setminus Q_{\ell,k}$ are irrelevant to the computation of the probability of disrupted link for sector $\ell$. Thus, it is clear to see that the term $\sum_{ K  \subset M \times M} \prod_{\{ (a,b) \in K^c \}} r_{a,b} \prod_{\{ (c,d) \in K \}} (1-r_{c,d})\phi_{i,\ell} (K,A)$ can be written as 
\begin{align*}
\Phi_{i,\ell}(A)= \mathbb{P} (\text{Some links among $|Q_k|$ links } (i,j)_{j \in Q_{\ell,k}} \text{ is disrupted }  )  \phi_{i,\ell} (.,A),
\end{align*}
where $\phi_{i,\ell} (.,A)$ is the value of $\phi_{i,\ell} (K,A)$ for some $K$ containing those disrupted link. Thus, 
$$u^r_i= \log(1-\rho)\displaystyle\sum_{\ell \in L}\Phi_{i,\ell} (A).$$

Now, let $m_k$ be the cardinality of $Q_k$ and let us consider a firm $i$ of input type $\bar \ell$ in cluster $Q_k$, i.e.  $i \in Q_{\bar\ell,k}$.

We remark that in both two disruption functions, $\phi_{i,\bar\ell} (A)=b_{\bar \ell,\bar \ell}$, which is independent of structure of $\mathcal{Q}$, so we focus on $\Phi_{i,\ell}(A)$ where $\ell \neq \bar \ell$. It is easy to see that
$$\Phi_{i,\ell}(A)=\sum_{q=1}^{m_k} \mathbb{P}(\text{$q$ links among $|Q_k|$ links } (i,j)_{j \in Q_{\ell,k}} \text{ are disrupted }  ) \phi_{i,\ell} (q,A),$$
where $ \phi_{i,\ell} (q,A)= \phi_{i,\ell}(K,A)$ for some $K$ such that $|K \cap \{(i,j)\mid j \in Q_{\ell,k}\}|=q$.
\medskip

If $\phi_i=\phi^{\min}_i$, we have $$\phi^{\min}_{i,\ell} (q,A)=\min_{j \in M_{\ell} \mid (i,j) \in K \text{ and }  a_{i,j} >0 } a_{i,j}=\frac{b_{\bar \ell,\ell}}{m_k}.$$
Thus,
\begin{align*}
\Phi^{\min}_{i,\ell}(A)&=\Big(1- \mathbb{P}(\text{All links } (i,j) \text{, where } j \in Q_{\ell,k}, \text{ are not disrupted}) \Big) \frac{b_{\bar \ell,\ell}}{m_k}\\
&=b_{\bar \ell, \ell}\dfrac{1-\displaystyle\prod_{j \in Q_{\ell,k}} (1-r_{i,j})}{m_k}.
\end{align*}
\medskip

If $\phi_i=\phi^{\text{sum}}_i$, we have $$\phi^{sum}_{i,\ell} (q,A)=\sum_{j \in M_{\ell} \mid (i,j) \in K \text{ and }  a_{i,j} >0} a_{i,j}=q\frac{b_{\bar\ell,\ell}}{m_k}.$$
Thus, $$\Phi^{sum}_{i,\ell}(A)=\frac{b_{\bar \ell,\ell}}{m_k}\sum_{q=1}^{m_k} q \times \mathbb{P}(\text{q links among $|Q_k|$ links } (i,j)_{j \in Q_{\ell,k}} \text{ are disrupted}).$$
The probability of exact $q$ disrupted links $(i,j_1)$,...,$(i,j_q)$, where $j_1,\ldots,j_q \in Q_{\ell,k}$, is
$$r_{i,j_1}\cdots r_{i,j_q}\prod_{\substack{j \neq  j_1,\ldots,j_q \\ j \in Q_{\ell,k}}}(1-r_{i,j}).$$ Applying Lemma \ref{lem_proba}, we have
$$\Phi^{sum}_{i,\ell}(A)=\frac{b_{\bar \ell,\ell}}{m_k}\sum_{j \in Q_{\ell,k}} r_{i,j}.$$
\medskip

Same computation as in the proof of Proposition \ref{prop_hick_neutral} gives that for all $i \in M_{\bar \ell}$,
\begin{align*}
\left((a_0)^T(I-A)^{-1}\right)_i=\frac{1}{n}\displaystyle\sum_{\ell \in L}a_{0,\ell}C_{\ell,\bar \ell}.
\end{align*}
As in the proof of Proposition \ref{poa-example}, we define $\sum_{\ell' \in L}a_{0,\ell'}C_{\ell',\bar\ell}= Q_{\bar\ell}$. Thus
\begin{align*}
(a_0)^T(I-A)^{-1}u^r&=\frac{1}{n}\displaystyle\sum_{\bar \ell \in L}Q_{\bar \ell}\Big( \sum_{i \in M_{\bar\ell}} u^r_i\Big)\\
&=\frac{1}{n}\log(1-\rho)\displaystyle\sum_{\bar \ell \in L}Q_{\bar \ell}\Bigg( \sum_{i \in M_{\bar\ell}} \bigg(\sum_{\ell \in L}\Phi_{i,\ell} (A)\bigg)\Bigg)\\
&=\frac{1}{n}\log(1-\rho)\displaystyle\sum_{\bar \ell \in L}Q_{\bar\ell}\Bigg( \sum_{\ell \in L} \bigg(\sum_{i \in M_{\bar\ell}}\Phi_{i,\ell} (A)\bigg)\Bigg).
\end{align*}

We then focus on the term $\displaystyle\sum_{i \in M_{\bar\ell}}\Phi_{i,\ell} (A)$, which is
\begin{equation}\label{sum of risk}
\displaystyle\sum_{k=1}^K \Big(\sum_{i \in Q_{\bar\ell,k}}\Phi_{i,\ell} (A)\Big) 
\end{equation}
If $\ell=\bar\ell$, then this sum equals to $n\times r_{i,i} \times b_{\bar\ell,\bar \ell}$ which is independent of the choice of $\mathcal{Q}$-cluster in either homogeneous risk or increasing risks. Therefore, the only concern arises when $\ell \neq \bar{\ell}$.

Now we are ready to prove the main result.

\begin{itemize}
\item[1)] If $\phi_i(K,A)=\phi^{\min}_i(K,A)$  and risk is homogeneous, we have $r_{i,j}=r$ for all links $(i,j)$ and $\Phi^{\min}_{i,\ell}(A)=b_{\bar \ell,\ell}\dfrac{1-(1-r)^{m_k}}{m_k}$. Thus,
\begin{equation*}
\displaystyle\sum_{k=1}^K \Big(\sum_{i \in Q_{\bar\ell,k}}\Phi^{\min}_{i,\ell} (A)\Big) =b_{\bar\ell,\ell}\displaystyle\sum_{k=1}^K \Big(1-(1-r)^{m_k}\Big).
\end{equation*}
Applying the inequality $(1-x)+(1-y)\geq 1-xy$ for all $0\leq x,y\leq 1$, one gets the highest value of $\displaystyle\sum_{i \in M_{\bar\ell}}\Phi_{i,\ell} (A)$ is when $|\mathcal{Q}|=1$ ($|\mathcal{Q}|=n$), which concludes our desired result.

\item[2)] If $\phi_i(K,A)=\phi^{\text{sum}}_i(K,A)$ and risk is homogeneous, we have $r_{i,j}=r$ for all links $(i,j)$ and $\Phi^{sum}_{i,\ell}(A)=b_{\bar \ell,\ell} \times r$. Our desired result is followed by
\begin{equation*}
\displaystyle\sum_{k=1}^K \Big(\sum_{i \in Q_{\bar\ell,k}}\Phi^{sum}_{i,\ell} (A)\Big) =n\times b_{\bar \ell,\ell}\times r.
\end{equation*}

\item[3)]  If $\phi_i(K,A)=\phi^{\text{sum}}_i(K,A)$ and risk increases with distance, we have
\begin{align*}
\displaystyle\sum_{k=1}^K \Big(\sum_{i \in Q_{\bar\ell,k}}\Phi^{sum}_{i,\ell} (A)\Big)& = 
b_{\bar\ell,\ell}\displaystyle\sum_{k=1}^K \Big(\dfrac{\sum_{i \in Q_{\bar\ell,k}}\sum_{j \in Q_{\ell,k}} r_{i,j}}{m_k}\Big)\\
& = 
b_{\bar\ell,\ell}\dfrac{r}{n}\left(n+\displaystyle\sum_{k=1}^K \Big(\dfrac{\sum_{i,j \in Q_k}|i-j|}{m_k}\Big)\right).
\end{align*}
Applying Lemma \ref{lem_sum_distance} several times, we can conclude that the maximal (minimal) value of $\displaystyle\sum_{i \in M_{\bar\ell}}\Phi_{i,\ell} (A)$ attains when $K=n$ (K=1).
\end{itemize}
\end{demo}

\begin{demo}[of Proposition \ref{prop-policy}]
The proof is straightforward by enumeration of the set of $\mathcal{Q}$-clustered equilibrium network. 
\end{demo} 

\section*{Appendix B: Extension to the CES case}

To simplify the exposition, let us assume the constant returns to scale approximation is valid and that profits can thus be approximated by revenues. Let us then consider the following economic structure.   The household only consumes  good $1.$ The producer of the consumption good $1$ uses good $A$ and good $B$ as inputs. 
The producer of good $B$ has a fixed production technology. The producer of good $A$ must strategically choose between two inputs: good $\gamma$ and good $\delta.$ We assume that both  $\gamma$ and  $\delta$ are produced by a linear production technology with labor as only input and that the production technology for $\gamma$ is more productive (and thus the production cost and the equilibrium price of $\gamma$ will always be lower).

At equilibrium, the producer of the consumption good will always receive a nominal demand equal to one (the total demand from the household assuming labor price is normalized to one). It will allocate this nominal demand between producers of $A$ and $B.$ The objective of $A$ (who is the only agent with an actual strategic choice to make) is to maximize the share of the nominal demand it receives.

First assume that the consumption good producer has a Leontieff production function. The share of nominal demand received by producer $A$ increases with the relative price of $A$ and $B.$ Producer $A$ thus has incentives to maximize its production cost and its price by purchasing input $\delta.$ 

Oppositely, if the consumption good producer has a linear production function, producer $A$ will receive a nominal demand only if it is (weakly) cheaper than producer $B.$ Producer $A$ thus has incentives to minimize its production cost and its price by purchasing input $\delta.$

\end{document}